\documentclass[twocolumn, aps, 10pt, pra, longbibliography,floatfix]{revtex4-2}

\usepackage{soul}
\usepackage{amsmath}   
\usepackage{amssymb}
\usepackage{mathtools}
\usepackage{graphicx}
\usepackage{dcolumn}
\usepackage{bm}
\usepackage{amsmath}
\usepackage{graphicx} 
\usepackage{amsfonts}
\usepackage{subfigure}
\usepackage{graphicx}
\usepackage{array}
\usepackage{color}
\usepackage{amssymb}%
\usepackage{makecell}
\usepackage[colorlinks=true,linkcolor=blue]{hyperref}%
\hypersetup{allcolors=blue}
\usepackage[normalem]{ulem}
\usepackage{xcolor}
\usepackage{tabularx,booktabs}
\usepackage{threeparttable}

\newcommand{\ket}[1]{\ensuremath{\left\vert #1 \right\rangle}}
\usepackage{mathtools}
\DeclarePairedDelimiterX\braket[2]{\langle}{\rangle}{#1 \delimsize\vert #2}

\newcolumntype{Y}{>{\centering\arraybackslash}X}

\begin{document}

\title{An optical atomic clock using \texorpdfstring{$4D_J$}{} states of rubidium}
\date{\today }

\author{A. Duspayev}
    \email{alisherd@umich.edu}
    \affiliation{Department of Physics, University of Michigan, Ann Arbor, MI 48109, USA} 
\author{C. Owens}
    \thanks{A.D. and C.O. contributed equally to this work.}
    \affiliation{Department of Physics, University of Michigan, Ann Arbor, MI 48109, USA} 
\author{B. Dash}
    \affiliation{Department of Physics, University of Michigan, Ann Arbor, MI 48109, USA}    
\author{G. Raithel}
    \affiliation{Department of Physics, University of Michigan, Ann Arbor, MI 48109, USA}

\begin{abstract}
We analyze an optical atomic clock using two-photon $5S_{1/2} \rightarrow 4D_J$ transitions in rubidium. Four one- and two-color excitation schemes to probe the fine-structure states $4D_{3/2}$ and $4D_{5/2}$ are considered in detail. We compare key characteristics of Rb $4D_J$ and $5D_{5/2}$ two-photon clocks. The $4D_J$ clock features a high signal-to-noise ratio due to two-photon decay at favorable wavelengths, low dc electric and magnetic susceptibilities, and minimal black-body shifts. Ac Stark shifts from the clock interrogation lasers are compensated by two-color Rabi-frequency matching. We identify a ``magic” wavelength near 1060~nm, which allows for in-trap, Doppler-free clock-transition interrogation with lattice-trapped cold atoms. From our analysis of clock statistics and systematics, we project a quantum-noise-limited relative clock stability at the $10^{-13}/\sqrt{\tau(s)}$-level, with integration time $\tau$ in seconds, and a relative accuracy of $\sim 10^{-13}$. We describe a potential architecture for implementing the proposed clock using a single telecom clock laser at 1550~nm, which is conducive to optical communication and long-distance clock comparisons. Our work could be of interest in efforts to realize small and portable Rb clocks and in high-precision measurements of atomic properties of Rb $4D_J$-states.
\end{abstract}

\maketitle
\section{Introduction}
\label{sec:intro}

 Recent efforts have led optical atomic clocks to be the most precise timekeeping devices, with many directions for further applications~\cite{atomicclocksreview, chargedclocksreview}. These include but are not limited to: the redefinition of the second~\cite{Dimarcq_2024}, tests of fundamental physics~\cite{reinhardt2007,blatt2008}, gravitational wave detection~\cite{kolkowitz2016} and searches for dark matter~\cite{derevianko2014, wcislo2018, kobayashi2022, filzinger2023}. The definition of the second~\cite{Leschiutta_2005} is currently based on the microwave Cs hyperfine transition measured in atomic fountain clocks~\cite{Heavner2014, Beattie2020}. These clocks utilize laser-cooled atoms and reach a fractional frequency stability below $10^{-15}$. Furthermore, the most precise optical atomic clocks can achieve stabilities on the order of $10^{-18}/\sqrt{\tau}$~\cite{kimprl2023} (with $\tau$ being the integration time in seconds), allowing for direct detection of gravitational red shifts with multiplexed atomic ensembles~\cite{bothwell2022,zheng2022}. Various atomic species, both neutral and charged, are being actively investigated as candidates for novel atomic-clock systems~\cite{huntemann2016,mcgrew2018,seiferle2018,holliman2022,zhiqiang2023}. 

Practical applications of atomic clocks, including geodesy and inertial navigation~\cite{Mehlstäubler_2018, ely2018, takamoto2022}, will generally benefit from a compact footprint, which is a challenge for the aforementioned best-performing atomic clocks. Alkali atoms remain relevant for this endeavor as various efforts are underway to ``package" the existing setups into portable devices~\cite{little2021, mcgilligan2022, martinez2023, bregazzi2023}. Microwave clocks based on the transition between hyperfine ground states in Rb are commonly utilized in commercial technology~\cite{camparo2007}. A relative stability reaching $4\times10^{-13}/\sqrt{\tau (s)}$ based on the optical two-photon $5S_{1/2}\rightarrow5D_{5/2}$ transition in Rb has been demonstrated in the context of realizing a portable optical atomic clock~\cite{martin2018}. The quadrupole transition in Cs, $6S_{1/2}\rightarrow5D_{5/2}$, at 685~nm has been proposed for a similar purpose~\cite{sharma2022analysis}. 

Here we analyze the optical two-photon $5S_{1/2} \rightarrow 4D_J$ transitions of Rb as a candidate for a portable and robust optical atomic clock. The $4D_J$ states in Rb are attractive for applications in modern quantum science and technology because two-photon transitions to these states are relatively strong and can be driven by readily available diode lasers with low to moderate output power~\cite{moon2009,wang2014,lim2022, Duspayev2023}. Further, the transitions into the $4D_J$ states via the Rb D$_{1}$- and D$_{2}$-lines involve telecom wavelengths for the upper stage ($\approx$~1476~nm and $\approx$~1529~nm, respectively). These can be used in quantum communication protocols~\cite{chanliere2006, huie2021}, as well as to network between distant optical atomic clocks for differential frequency comparisons~\cite{komar2016, BACON2021, nichol2022, eustice2023}. Moreover, the $4D_J$ states can be utilized in Rydberg physics applications such as electric field sensing using high-angular-momentum Rydberg states~\cite{duspayev2023highell} and all-optical preparation of Rydberg molecules~\cite{shafferreview,duspayev2021} and circular Rydberg atoms~\cite{cardman2020}. 

Our paper is structured as follows. In Sec.~\ref{sec:general} we 
discuss general aspects of the proposed Rb $4D_J$ clock. Considerations on level structure, fluorescence decay channels, and line pulling from transitions with non-vanishing first-order Zeeman shifts lead into our selection of four specific clock modes. 
In Sec.~\ref{sec:scheme} the modes are discussed in detail and ranked by promise, with an emphasis on ac-shift cancellation, number of laser sources required and corresponding beam powers, and fluorescence detection efficiency. 
In Secs.~\ref{sec:statistics} and~\ref{sec:syst} we evaluate statistical and systematic uncertainties of the clock frequency, respectively, and summarize key systematics.  In Sec.~\ref{sec:disc} we discuss selected aspects, present a possible clock implementation and conclude the paper.

\begin{figure*}[t!]
 \centering
  \includegraphics[width=0.99\textwidth]{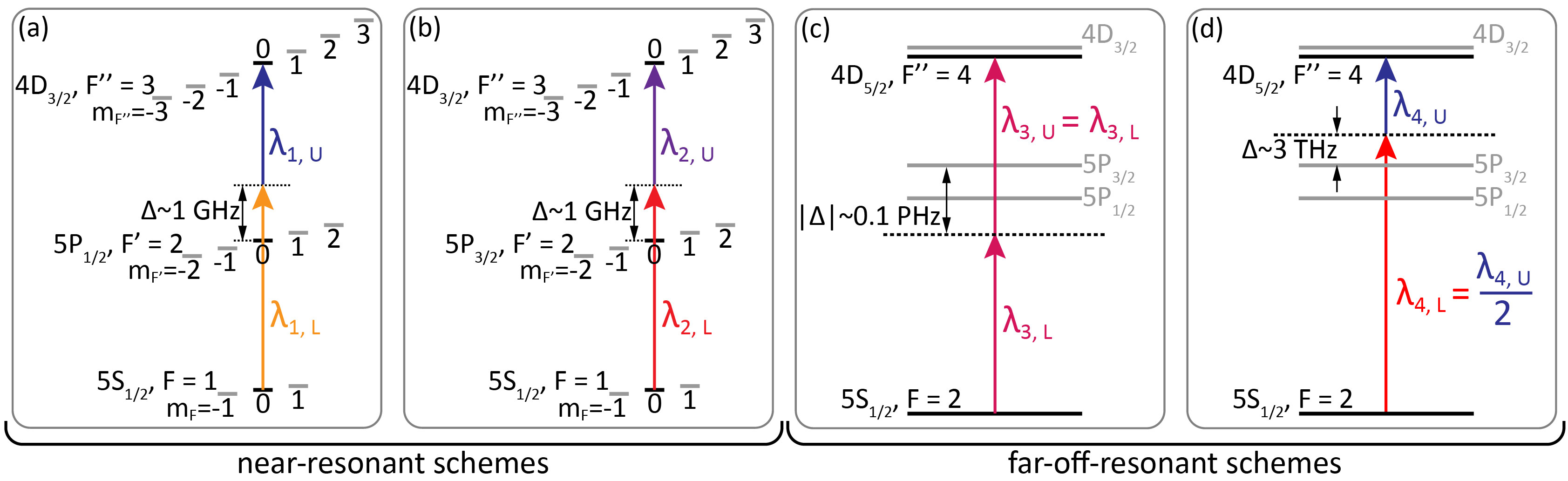}
  \caption{Energy level diagrams for the considered $^{87}$Rb $4D_J$ clock designs (not to scale). (a) and (b) are $4D_{3/2}$-schemes with excitation via the $D_1$ and $D_2$ lines, respectively. (c) and (d) show two one-color and two-color two-photon far-off-resonant $4D_{5/2}$-schemes, respectively. Most hyperfine and magnetic sub-structures are omitted in (c) and (d); $m_F = m_{F''} = 0$ also applies there. See text for details.} 
  \label{fig:scheme}
\end{figure*}

\section{General concepts}
\label{sec:general}

\subsection{Overview}

The clock schemes under consideration involve two-photon $5S_{1/2} \rightarrow 4D_J$ transitions in $^{87}$Rb. We discuss several schemes, depicted in Fig.~\ref{fig:scheme}, that primarily differ in the detunings relative to the intermediate $5P_J$-states, the transition detection methods, the severity of ac level shifts caused by the excitation lasers, and the Doppler shifts present. 
In the schemes in Figs.~\ref{fig:scheme} (a) and (b), the $4D_{3/2}$ state is utilized as the upper clock state, the two-photon excitation proceeds relatively close to resonance through one of the two $5P_J$ states, and the clock transition is monitored by detecting the fluorescence from decay through the other $5P_J$ state. In the schemes in Figs.~\ref{fig:scheme} (c) and (d), we utilize far-off-resonant excitations into the $4D_{5/2}$ state and detection of fluorescence from decay through the $5P_{3/2}$-state. We discuss advantages and drawbacks of the schemes, and compare aspects of the $4D_J$ and the more commonly-used $5D_{5/2}$-clocks~\cite{martin2018}. Throughout our paper, we use the notation that hyperfine quantum numbers with no, one and two primes refer to lower-, intermediate- and upper-state levels, respectively.

\subsection{Line pulling due to first-order Zeeman effect}
\label{subsec:zeeman1st}

The atomic clock frequency is the sum of the frequencies of lower- and upper-transition lasers locked to the desired $5S_{1/2}$ to $4D_J$ two-photon transition. 
For a relative clock uncertainty of $1 \times 10^{-13}$, the uncertainty of the difference between the center values of the $5S_{1/2}$ and $4D_{J}$ energy levels must not exceed $\sim h \times 60$~Hz. This necessitates near-complete elimination of the effects of first-order Zeeman shifts and suppression of the remaining quadratic Zeeman shifts from a bias magnetic field, $B_{bias}$, which is applied to define a quantization axis. Since the $4D_J$ decay rate is $\Gamma_{4D} \approx 2 \pi \times 2$~MHz, a bias magnetic field $B_{bias} \gtrsim 5~$G, would be necessary to isolate a single Zeeman component of the clock transition with vanishing first-order Zeeman shift. Such a large bias field is deemed prohibitive because of the incurred second-order Zeeman shifts (see Sec.~\ref{subsec:bfield}). Here, we consider bias magnetic fields $B_{bias} \lesssim 100$~mG. Magnetic shifts, as well as other systematic shifts, are then due to unwanted, weak perturber lines that are hidden underneath the targeted clock-transition line and that slightly pull the line center.

We consider a set of $i=1, ..., i_{max}$ spectral lines with relative line strengths $p_i$ and detunings $\delta_i$, with $\sum_i p_i = 1$. Typically, there is a desired, main Zeeman line with a near-zero  $\delta_{i0}$ and near-unity $p_{i0}$. The main line is pulled by weak Zeeman and other perturber lines that have $\delta_i \ll \Gamma_{4D}$ and small $p_i$. Considering a symmetric homogeneous line shape, which could be a Lorentzian, a saturated Lorentzian, etc., it is easy to show that the observed shift of the line center, $\delta$, follows the intuitive equation 
\begin{equation} 
\delta = \sum_i \delta_i p_i \quad .
\label{eq:pull}
\end{equation}

The Zeeman components of the $\ket{5S_{1/2}, F} \rightarrow \ket{4D_J, F''}$ clock line are characterized by Zeeman shifts $\delta_i(m_F, m_{F''})$ that are dependent on the initial- and final-state magnetic quantum numbers, $m_F$ and $m_{F''}$, atomic line strengths $W(m_F, m_{F''})$ that are dependent on invariable atomic electric-dipole matrix elements, clock-laser polarizations and intermediate-state detunings, and on initial-state probabilities $P(m_F)$ that reflect the magnetization state of the atom sample in the $5S_{1/2}$ ground state. Then $p_i$ in Eq.~\ref{eq:pull} is given by $p_i = P(m_F) W(m_F, m_{F''})$, with proper normalization $\sum_i p_i = 1$.  

The assumed bias field $B_{bias} \lesssim 100$~mG gives rise to $\delta_i(m_F, m_{F''})$-values in the range of $2 \pi \times $100~kHz. For a relative clock uncertainty of $10^{-13}$, the line-pulling resultant from Eq.~\ref{eq:pull} must then satisfy $\vert \delta \vert \lesssim 2 \pi \times 60~$Hz. Practical solutions include unmagnetized atom samples with vanishing stray magnetization and $\pi$-polarized clock lasers, or samples prepared by high-fidelity optical pumping into a magnetic ground-state level with $m_F=0$. In the former case, it is $P(m_F) \approx 1/(2F+1)$ for all $m_F$ and $\sum_{m_F} P(m_F) m_F \approx 0$, {\sl{i.e.}}, the Zeeman lines are symmetric about the line center. In the latter case, it is $P(m_F) \lesssim 1 $ for $m_F=0$, $P(m_F) \sim 0$ for $m_F \ne 0$, and $\sum_{m_F} P(m_F) m_F \approx 0$. Cases other than these two may fail due to line pulling from asymmetrically-placed perturber lines with large linear Zeeman shifts.

For specificity, here we mostly consider clock schemes in which the upper and lower states have magnetic quantum numbers $m_F = m_{F''}=0$, eliminating linear Zeeman shifts of the clock transition and leaving only a weak quadratic Zeeman shift to contend with. To drive two-photon transitions between states with $m_{F} = m_{F''}=0$, one may employ clock-drive lasers that are both $\pi$-polarized  ($\Delta m = 0$), or that are $\sigma$-polarized with opposite helicity ($\Delta m = \pm 1$). 
We select $\pi$-polarized clock-drive lasers because linearly polarized light is less susceptible to polarization errors than circularly polarized light. Polarization errors must be minimized because they would result in
weak $\Delta m \ne 0$ perturber lines with linear Zeeman shifts, which would likely cause line pulling $ \vert \delta \vert> 2 \pi \times 60$~Hz, as explained above. 
Even for clean $\pi$-polarizations, the anomalous Zeeman effect results in $m_F$-dependent linear Zeeman shifts of the $m_F \rightarrow m_{F''}=m_F$ clock transitions. Anomalous Zeeman shifts range between 350~kHz/G and 1.26~MHz/G for the clock modes in Fig.~\ref{fig:scheme}. To limit line pulling from $m_F \ne 0$ perturber lines, the optical pumping into $m_F = 0$ must be efficient, and spurious $m_F \ne 0$ populations must be symmetrically distributed about $m_F = 0$. For $B_{bias} \lesssim 100$~mG we expect to be able to meet the condition $\vert \delta \vert \lesssim 2 \pi \times 60$~Hz with light-polarization  and optical-pumping inefficiencies in the sub-percent range. 

Polarization errors must also be avoided because they would cause Rabi-frequency fluctuations of the clock transitions. Such fluctuations would be detrimental to ac-shift cancellation via Rabi-frequency matching between lower and upper clock transitions, which is employed to reduce clock-laser-induced ac-shifts (see Secs.~\ref{subsec:acshift} and~\ref{sec:scheme}).

\subsection{Line pulling from off-resonant \texorpdfstring{$4D_J$}{} hyperfine levels}
\label{subsec:pulling}

The hyperfine splittings of $4D_J$ are sub-100-MHz and are larger in $^{87}$Rb than in $^{85}$Rb by about a factor of three. In order to minimize the effects of line pulling from off-resonant $4D_J$ hyperfine 
lines, we select the hyperfine levels $F''=3$ of $^{87}$Rb $4D_{3/2}$ for the near-resonant clock schemes in Figs.~\ref{fig:scheme} (a) and (b), and the level $F''=4$ of $^{87}$Rb $4D_{5/2}$ for the far-off-resonant clock schemes
in Figs.~\ref{fig:scheme} (c) and (d). These hyperfine lines exhibit maximal separations from other $4D_J$ hyperfine lines.
Maximizing the $4D_J$ hyperfine separation also reduces second-order Zeeman shifts (see Sec.~\ref{subsec:bfield}).

\subsection{Selection of specific clock transitions}
\label{subsec:select}

To achieve a high signal-to-noise ratio (SNR) of the detected $4D_J$-fluorescence, dichroic optics and spectral filters must be employed to eliminate scattered drive-laser stray light from the fluorescence detector.
Two-photon excitation of the Rb $4D_J$ states proceeds via the intermediate $5P_J$ states, which also are the only intermediate states through which the atoms decay back into the ground state. In two of the four schemes discussed [Figs.~\ref{fig:scheme} (a) and (b)], the $4D_J$-excitation is fairly close to resonance with one of the intermediate $5P_J$-states. The spectral filters transmit fluorescence from decays through the other $5P_J$-state. This forces the use of $4D_{3/2}$ as the upper clock state in Figs.~\ref{fig:scheme} (a) and (b). 

For $m_F = m_{F''}=0$, the $\pi$-couplings follow the selection rules $F \ne F'$ and $F' \ne F''$. This simplifies the relations between Rabi frequencies, detunings, clock fluorescence rates and ac shifts because there is only one intermediate level and only one intermediate detuning, $\Delta$. The near-resonant two-color excitation schemes in Fig.~\ref{fig:scheme} (a) and (b) are
\begin{widetext}
\begin{subequations} 
\begin{eqnarray}
\vert 5S_{1/2}, F=1, m_F=0  \rangle 
&
\xrightarrow{\lambda_{1,L}}
& 
\vert 5P_{1/2}, F'=2, m_{F'}=0 \rangle 
\xrightarrow{\lambda_{1,U}}
\vert 4D_{3/2}, F''=3, m_{F''}=0 \rangle  \\
\vert 5S_{1/2}, F=1, m_F=0 \rangle 
&
\xrightarrow{\lambda_{2,L}}
&
\vert 5P_{3/2}, F'=2, m_{F'}=0 \rangle 
\xrightarrow{\lambda_{2,U}} 
\vert 4D_{3/2}, F''=3, m_{F''}=0 \rangle \quad,
\label{eq:mode12}
\end{eqnarray}
\end{subequations} 
\end{widetext}
respectively. Here, the transition wavelengths $\lambda$ carry subscripts 1 and 2 for excitation via the $D_1$ and $D_2$ lines, and $U$ and $L$ for the respective lower and upper transitions. Optical pumping into the lower clock state $\vert 5S_{1/2}, F=1,  m_F=0 \rangle$ is performed with an auxiliary $\pi$-polarized, low-power laser beam resonant with the $\vert 5S_{1/2}, F=1 \rangle  \rightarrow \vert 5P_J, F'=1 \rangle$ transition, with an addition of a weak $x$-polarized clock re-pumper beam on $\vert 5S_{1/2}, F=2 \rangle  \rightarrow \vert 5P_J, F'=2 \rangle$.

In Figs.~\ref{fig:scheme} (c) and (d) we show the two far-off-resonant drive schemes considered. For those, the lower excitation wavelengths are sufficiently far away from both the $D_1$ and $D_2$ lines such that decays through both $5P_J$-states can be simultaneously detected. An efficient scheme utilizes two-photon $\pi$-polarized ($\Delta m =0$) drives into $4D_{5/2}$,
\begin{equation}
\vert 5S_{1/2}, F=2,  m_F=0 \rangle  \rightarrow \vert 4D_{5/2}, F''=4,  m_{F''}=0 \rangle 
\label{eq:mode34}
\quad \end{equation} 
in $^{87}$Rb. This transition is closed with regard to $F$ and $F''$.  In those schemes, optical pumping into the lower clock state is performed by weak $\pi$-polarized laser beams resonant with the $\vert 5S_{1/2}, F=2 \rangle  \rightarrow \vert 5P_J, F'=2 \rangle$ transition, plus a weak clock re-pumper beam on $\vert 5S_{1/2}, F=1 \rangle  \rightarrow \vert 5P_J, F'=2 \rangle$.

It is noted that the clock-laser wavelengths are $\in [$774~nm,~795~nm$]$, near 1033~nm, or 
$\in [$1476~nm,~1550~nm$]$. The latter interval is in the S- and C-bands of telecommunications. Narrow-line lasers at these wavelengths are readily available. In most cases, the powers required are in the range of tens to a few hundred mW. For the near-resonant schemes in Figs.~\ref{fig:scheme} (a) and (b) two excitation lasers are required,  while for the far-off-resonant schemes in Figs.~\ref{fig:scheme} (c) and (d) only a single laser source is needed.

\subsection{Fluorescence detection}
\label{subsec:detectors}

The $4D_J$-fluorescence has a yield of two photons per atom in two optical bands that both allow efficient photo-detection, which is conducive to a high SNR of the measured clock fluorescence. The only four decay wavelengths are 795~nm, 780~nm, 1476~nm and 1529~nm, for which we can leverage a range of well-developed and affordable photodetectors. Germanium and InGaAs photodiodes have good quantum efficiencies $\gtrsim 70\%$ and can be moderately cooled with one- or two-stage thermo-electric coolers to reduce thermal background currents. 
Ge sensors could be preferable because they are available with large sensitive areas, as required for large solid angles in fluorescence detection. To measure the 780-nm and 795-nm fluorescence, large-area Si diodes may be used, which also offer high efficiency. In all cases, to achieve a high SNR, dichroic optics and optical filters are employed to reduce optical noise caused by detection of ambient background light and scattered light from the clock excitation lasers. 

The described fluorescence measurement schemes for Rb $4D_J$ clocks compare favorably well with fluorescence measurement in Rb $5D_J$ clocks. In the latter, fluorescence is typically measured on the $6P_J$ to $5S_{1/2}$ decay channel near 420~nm~\cite{martin2018, martinpra2019}. This decay channel has a yield of only about one blue photon for every four $5D_J$-atoms. Moreover, blue-light photodetectors typically have quantum efficiencies $\lesssim 35\%$. 

\subsection{ac shift cancellation}
\label{subsec:acshift}

Ac shifts from the lower and upper clock transitions are in the $\gtrsim 10$-kHz range. Fortunately, lower and upper clock states experience ac shifts in the same direction. 
If two separate laser beams are applied to drive the lower and upper clock transitions, as in the schemes discussed in Secs.~\ref{subsec:mode1}, \ref{subsec:mode2} and~\ref{subsec:mode4}, separate intensity controls of the two beams allow for cancellation of the net clock-laser-induced ac shift of the clock transition.

Ac shift cancellation is not possible with single-color two-photon excitation. In single-color two-photon Rb $4D_{5/2}$ and $5D_{5/2}$~\cite{martin2018, gerginov2018, perrella2019} clocks, discussed in Secs.~\ref{subsec:mode3} and \ref{subsec:mode5}, the ac shift typically is on the order of tens of kHz and cannot be cancelled, leaving intensity variations of the excitation laser as a limiting factor in the clock uncertainty.

\section{Detailed discussion of specific clock drive modes}
\label{sec:scheme}

\subsection{Near-resonant \texorpdfstring{$5S_{1/2}$ - $5P_{1/2}$ -$4D_{3/2}$}{} two-color drive}
\label{subsec:mode1}

We first discuss the case of two $\pi$-polarized excitation fields at $\lambda_{1,L} = 794.96$~nm and $\lambda_{1,U} = 1475.64$~nm that drive the $\vert 5S_{1/2}, F=1, m_F=0 \rangle$ $\rightarrow$ $\vert 5P_{1/2}, F'=2, m_{F'}=0 \rangle$ and $\vert 5P_{1/2}, F'=2, m_{F'}=0 \rangle$ $\rightarrow$ $\vert 4D_{3/2}, F''=3, m_{F''}=0 \rangle$ transitions of $^{87}$Rb [see Fig.~\ref{fig:scheme}~(a)]. The respective Rabi frequencies 
are denoted $\Omega_{SP}$ and $\Omega_{PD}$. Since selection rules only allow the intermediate state $F'=2$, the intermediate detuning  $\Delta$ is well-defined, and the decay rate out of the $4D_{3/2}$ level is, in the applicable case of low saturation,
\begin{equation}
\gamma_{4D} = \frac{\Omega_{SD}^2}{\Gamma_{4D}} = \frac{\Omega_{SP}^2 \Omega_{PD}^2}{4 \Delta^2 \Gamma_{4D}} \quad,
\label{eq:g4D}
\end{equation}
where the two-photon Rabi frequency $\Omega_{SD} = \Omega_{SP} \Omega_{PD} / (2 \Delta)$ and the $4D_{3/2}$ natural decay rate $\Gamma_{4D} = 2 \pi \times 1.92$~MHz. By comparison, unwanted off-resonant photon scattering from the intermediate level occurs at a rate of
\begin{equation}
\gamma_{5P} = \frac{\Omega_{SP}^2 \Gamma_{5P}} {4 \Delta^2} \quad, 
\label{eq:g5P}
\end{equation}
with the $5P_{1/2}$ natural decay rate $\Gamma_{5P} = 2 \pi \times 5.746$~MHz. It is desired to minimize background scatter and to avoid atom heating (see below). Hence, we are aiming for a large ratio of beneficial photon scattering over unwanted one, 
\begin{equation}
\frac{\gamma_{4D}}{\gamma_{5P}} = \frac{\Omega_{PD}^2}{\Gamma_{5P} \Gamma_{4D}} \quad.
\label{eq:gratio}
\end{equation}
Note the $\Delta$-independence of this ratio. The only adjustable variable in this ratio is the upper-transition Rabi frequency $\Omega_{PD}$, which one will want to choose sufficiently large. 

The light shifts of the clock levels can be separated into near-resonant terms from the clock transitions and terms from far-off-resonance atomic levels. For the case of near-resonant clocks, the former are highly dominant and are given by $\Omega_{SP}^2/ (4 \Delta)$ and $\Omega_{PD}^2/ (4 \Delta)$ for the respective $\vert 5S_{1/2}, F=1, m_F=0 \rangle$ and $\vert 4D_{3/2}, F''=3, m_{F''}=0 \rangle$ clock levels. Here, we desire that the near-resonant ac shifts of the lower and upper clock states will be approximately matched, so that the clock-laser-induced ac shift of the transition frequency is approximately cancelled out. The cancellation is accomplished    
by adjusting the lower- and upper-transition laser intensities so that $\vert \Omega_{SP}^2 - \Omega_{PD}^2 \vert < \epsilon \Omega^2$, with an experimental imbalance parameter $\epsilon \ll 1$ and $\Omega^2 = (\Omega_{SP}^2 + \Omega_{PD}^2)/2$. The residual ac shift of the clock transition due to the near-resonant $5P_{1/2}$-state then is 
\begin{equation}
\vert \delta \omega_{ac} \vert \approx \frac{\epsilon \Omega^2}{4 \Delta} \quad.
\label{eq:acnear1}
\end{equation}

For a meaningful comparison of clock drive modes, we set $\gamma_{4D} = 10^3$~s$^{-1}$ per atom in all drive modes considered. This value suffices to reach the SNR of $10^4$ as required in Sec.~\ref{sec:statistics}. With given $\gamma_{4D}$, it is then found that the near-resonant ac shift of the clock-transition angular frequency follows 
\begin{equation}
\vert \delta \omega_{ac} \vert \approx 
\epsilon \sqrt{\gamma_{4D} \Gamma_{4D}}/2 \quad,
\label{eq:acnear2}
\end{equation}
where the result is in units of rad/s, and the rates under the square root are entered in units of 1/s (as provided above). The clock shift in Eq.~\ref{eq:acnear2} solely depends on the experimental imbalance parameter $\epsilon$, the desired $4D$ photon scattering rate $\gamma_{4D}$, and the natural $4D_{3/2}$ decay rate, $\Gamma_{4D}$, while $\Delta$ and the Rabi frequencies $\Omega_{SP} \approx \Omega_{PD}$ drop out. This occurs under the provision that $\Delta \gg \Gamma_{5P}$. There is, however, an incentive to keep $\Delta$ below certain bounds because the intensities and powers of both drive beams increase linearly in $\Delta$, and because the far-off-resonant light shifts increase linearly with the drive intensities. 

In the following example, we use $\gamma_{4D} = 10^3$~s$^{-1}$, $\Delta = 2 \pi \times 1$~GHz, and $\epsilon = 0.5 \%$. From Eq.~\ref{eq:g4D} one finds matched lower and upper Rabi frequencies $\Omega \approx \Omega_{SP} \approx \Omega_{PD} \approx 2 \pi \times 5.91~$MHz. The ratio in Eq.~\ref{eq:gratio} then is 3.17, which is fairly favorable. From the electric-dipole moments for the lower and upper transitions,
\begin{widetext}
\begin{eqnarray}
\langle 5S_{1/2}, F=1, m_F=0 \vert e \hat{z} \vert 5P_{1/2}, F'=2, m_{F'}=0 \rangle & = &  1.72~e a_0 \nonumber \\
\langle 5P_{1/2}, F'=2, m_{F'}=0 \vert e \hat{z} \vert 4D_{3/2}, F''=3, m_{F''}=0 \rangle & = &  3.11~e a_0 \quad,
\label{eq:mat1}
\end{eqnarray}
\end{widetext}
one finds the respective laser electric fields, intensities, and beam powers for given beam sizes. For instance, for Gaussian beams with equal beam waist parameters $w_0 = 1$~mm one finds lower- and upper-transition beam powers of only about 150~$\mu$W and 50~$\mu$W. For $\epsilon = 0.5\%$, Eqs.~\ref{eq:acnear1} and~\ref{eq:acnear2} yield an imbalance of ac-shifts due to the $5P_{1/2}$-state of about $2 \pi \times$ 40~Hz, which is below the limit of $2 \pi \times$ 60~Hz set in Sec.~\ref{sec:general}.

In order to aanalyze the background ac shift from far-off-resonant atomic states, we compute off-resonant polarizabilities as described in \cite{cardman2021, atoms10040117}. For electric-dipole matrix elements of transitions between lower-lying atomic states we use values provided in~\cite{safronova2004, safronova2011}. 
Matrix elements for transitions into higher-lying states are computed with our own codes~\cite{reinhardtA2007}, which utilize model potentials from~\cite{Marinescu1994}. For the case in this Section, the background ac shifts are computed by summing over all electric-dipole-coupled perturbing states, but excluding the shift from the separately-treated near-resonant state $5P_{1/2}$. The far-off-resonant polarizabilities are, in atomic units, 5512 and 428 for $\vert 5S_{1/2}, * \rangle$ in laser fields of $\lambda_{1,L} = 794.96$~nm and $\lambda_{1,U} = 1475.64$~nm wavelengths, respectively, and 982 and 4140 for $\vert 4D_{3/2}, F''=3, m_{F''}=0 \rangle$ in the same respective fields. The polarizability uncertainties are estimated at $1\%$, based on uncertainties of the matrix elements used. The resultant ac shift of the clock transition due to far-off-resonant atomic states is about $ 2 \pi \times$~3~Hz, which is well below the limit of $2 \pi \times$ 60~Hz set in Sec.~\ref{sec:general}. In the presented model, $\Delta$ has an allowable upper limit because, under the constraint of a fixed $\gamma_{4D}$, upper- and lower-transition intensities scale linearly in $\Delta$ (see Eq.~\ref{eq:g4D}). In the present case, an increase in $\Delta$ from $2 \pi \times$~1~GHz to about $2 \pi \times$~20~GHz would result in a clock-transition shift of $\sim 2 \pi \times$~60~Hz, the limit set in Sec.~\ref{sec:general}. 

We lastly consider the detection of $4D_{3/2}$ clock fluorescence for the $5S_{1/2}$ - $5P_{1/2}$ -$4D_{3/2}$ drive mode. The branching ratio of the $4D_{3/2}$ decay is about 16$\%$ through $5P_{3/2}$ versus 84$\%$ through $5P_{1/2}$. We assume that both excitation wavelengths, and with it any $4D_{3/2}$ clock fluorescence through the $D_1$ line, will have to be filtered out before photo-detection. Hence, only about 1 out of 6 decays can potentially be detected. We therefore consider the near-resonant $5S_{1/2}$ - $5P_{1/2}$ -$4D_{3/2}$ clock drive mode to be less competitive than the drive modes discussed next.   

\subsection{Near-resonant \texorpdfstring{$5S_{1/2}$ - $5P_{3/2}$ -$4D_{3/2}$}{} two-color drive}
\label{subsec:mode2}

Here, we discuss the case of two $\pi$-polarized excitation fields at
$\lambda_{2,L} = 780.24$~nm and $\lambda_{2,U} =1529.26$~nm that drive the $\vert 5S_{1/2}, F=1, m_F=0 \rangle$ $\rightarrow$ $\vert 5P_{3/2}, F'=2, m_{F'}=0 \rangle$ and $\vert 5P_{3/2}, F'=2, m_{F'}=0 \rangle$ $\rightarrow$ $\vert 4D_{3/2}, F''=3, m_{F''}=0 \rangle$ transitions [see Fig.~\ref{fig:scheme}~(b)]. The analysis given in Sec.~\ref{subsec:mode1} carries over, with the replacement 
\begin{widetext}
\begin{eqnarray}
\langle 5S_{1/2}, F=1, m_F=0 \vert e \hat{z} \vert 5P_{3/2}, F'=2, m_{F'}=0 \rangle & = &  1.73~e a_0 \nonumber \\
\langle 5P_{3/2}, F'=2, m_{F'}=0 \vert e \hat{z} \vert 4D_{3/2}, F''=3, m_{F''}=0 \rangle & = &  0.628~e a_0 \quad.
\label{eq:mat2}
\end{eqnarray}
\end{widetext}
For same parameters as in Sec.~\ref{subsec:mode1}, namely $\gamma_{4D} = 10^3$~s$^{-1}$, $\Delta = 2 \pi \times 1$~GHz, and $\epsilon = 0.5 \%$, the laser electric field, intensity  and beam power for the upper transition are larger due to the smaller upper-transition matrix element in Eq.~\ref{eq:mat2}. For instance, for Gaussian beams with $w_0 = 1$~mm one finds lower- and upper-transition beam powers of about 150~$\mu$W and 1~mW. The near-resonant ac-shift imbalance from Eqs.~\ref{eq:acnear1} and~\ref{eq:acnear2} remains at about $2 \pi \times$ 40~Hz.

The far-off-resonant ac polarizabilities from perturbing states excluding $5P_{3/2}$, calculated as described in Sec.~\ref{subsec:mode1}, are, in atomic units, -2714 and 417 for $\vert 5S_{1/2}, * \rangle$ in fields of $\lambda_{2,L} =780.241$~nm and $\lambda_{2,U} =1529.26$~nm wavelengths, respectively, and 
-368 and -6316 for $\vert 4D_{3/2}, F''=3, m_{F''}=0 \rangle$ in the same respective fields.
The magnitude of the net ac shift of the clock transition due to far-off-resonant atomic states is about $ 2 \pi \times$ 24~Hz, which is below the limit of $2 \pi \times$ 60~Hz set in Sec.~\ref{sec:general}. However, as a result of the larger upper-transition intensity, $\Delta$ has a lower allowed upper limit than in Sec.~\ref{subsec:mode1}, namely about $ 2 \pi \times$ 2~GHz.

For a clock drive through the $D_2$ line, the ratio in Eq.~\ref{eq:gratio} is 3.04, which is still fairly favorable. In the fluorescence detection, we filter out decays through $5P_{3/2}$ and only detect decays 
through $5P_{1/2}$. Since the branching ratio of the $4D_{3/2}$ decay favors decay through $5P_{1/2}$ over decay through $5P_{3/2}$ by about a factor of 5, in this clock drive mode 5 out of 6 decays are detectable. We therefore consider the near-resonant $5S_{1/2}$ - $5P_{3/2}$ -$4D_{3/2}$ clock mode to be quite competitive.

\subsection{Far-off-resonant \texorpdfstring{$5S_{1/2}$ -$4D_{5/2}$}{} Doppler-free single-color two-photon drive}
\label{subsec:mode3}

The concept of Doppler-free, single-color two-photon spectroscopy with counter-propagating beams can be extended from the Rb $5D_{5/2}$ clock~\cite{martin2018} to Rb $4D_{5/2}$~\cite{Roy2017}. While the laser wavelength of $\lambda_{3,*} =1033.314$~nm [see Fig.~\ref{fig:scheme}~(c)] is quite far-off-resonant from the intermediate $5P_{3/2}$ state, the absence of any other intermediate states between $5S_{1/2}$ and $4D_{5/2}$ as well as the dominant size of the transition matrix elements through $5P_{3/2}$~\cite{safronova2004, safronova2011} make some of the equations from Sec.~\ref{subsec:mode1} applicable to this clock mode.
The main difference relies in the fact that the transition Rabi frequencies $\Omega_{SP}$ and $\Omega_{PD}$ cannot be matched because the Doppler-free two-photon method employs laser beams
of exactly the same frequency for lower and upper clock transitions. The Rabi-frequency ratio equals that of the dipole matrix elements,  
\begin{widetext}
\begin{eqnarray}
\langle 5S_{1/2}, F=2, m_F=0 \vert e \hat{z} \vert 5P_{3/2}, F'=3, m_{F'}=0 \rangle & = &  2.32~e a_0 \nonumber \\
\langle 5P_{3/2}, F'=3, m_{F'}=0 \vert e \hat{z} \vert 4D_{5/2}, F''=4, m_{F''}=0 \rangle & = &  3.36~e a_0 \quad.
\label{eq:mat3}
\end{eqnarray}
\end{widetext}
As a result, the ac shifts from the drive beams cannot be cancelled. Favorable characteristics of this clock mode 
include that it can be applied in Rb vapor cells~\cite{Roy2017}, due to its Doppler-free character. Further, drive and fluorescence wavelengths are well-separated, the drive is on a cycling transition with regard to the relevant $F$ and $F''$-values, and there is a yield of two detectable photons for each $4D_{5/2}$ atom.

The value of $\Delta$ is fixed at  $\Delta= - 2 \pi \times 9.41 \times 10^4$~GHz. Requiring the same $\gamma_{4D} = 10^3$~s$^{-1}$ as in Secs.~\ref{subsec:mode1} and~\ref{subsec:mode2} and using Eq.~\ref{eq:g4D} (for $\Omega_{SP}$ and $\Omega_{PD}$ that are not equal but fixed in ratio using values in Eq.~\ref{eq:mat3}), one finds  a very high drive-laser intensity of $3.3 \times 10^6$~W/m$^2$. This results in uncomfortably high beam powers. For Gaussian beams with waist parameter $w_0=1$~mm, one would require 5.2~W per beam. Due to the large value of $\vert \Delta \vert$, the ratio $\gamma_{4D}/\gamma_{5P} = 4.3 \times 10^5$, which is very favorable. 

Since this clock mode is far-off-resonant from any intermediate levels, there is no advantage in distinguishing between near-resonant and far-off-resonant ac shifts. The ac polarizabilities of 
$\vert 5S_{1/2}, * \rangle$ and $\vert 4D_{5/2}, F''=4,
m_{F''}=0 \rangle$ at $\lambda_{3,*} =1033.314$~nm, summed over all coupled intermediate states (including $5P_{3/2}$), are 726 and 1745 in atomic units, respectively. For the drive-laser intensity stated in the previous paragraph, one finds respective ac shifts of $- 2 \pi \times$~11.3~kHz and $- 2 \pi \times$~27.0~kHz. The differential shift for the clock transition of  $- 2 \pi \times$~15.7~kHz exceeds the magnitude-limit of $2 \pi \times$~60~Hz set in Sec.~\ref{sec:general} by a factor of about 250. In laboratory experiments aimed at measuring hyperfine structures~\cite{Roy2017} and other atomic properties, the ac-shift problem may be ameliorated by extrapolating the line positions to zero drive power~\cite{Duspayev2023}. However, in a clock application one would have to compromise between ac clock shifts and clock scattering rates $\gamma_{4D}$, forcing a low $\gamma_{4D}$. A low $\gamma_{4D}$ results, in turn, in a low clock interrogation bandwidth and SNR.

Overall, we believe that the 1033-nm far-off-resonant $5S_{1/2}$ -$4D_{5/2}$ clock is less competitive than other schemes because of uncompensated ac shifts, the high laser-power requirement, and low bandwidth and SNR.

\subsection{Far-off-resonant \texorpdfstring{$5S_{1/2}$ -$4D_{5/2}$}{} two-color drive}
\label{subsec:mode4}

The intermediate state $5P_{3/2}$ splits the energy gap between the $5S_{1/2}$ and $4D_{5/2}$ clock states into two segments with a ratio of about 2 to 1. Hence, a single laser source at $\lambda_{4,U} =1549.971$~nm and its second harmonic at
$\lambda_{4,L} =774.985$~nm can be used to realize a single-laser, two-color, far-off-resonant $5S_{1/2}$ -$4D_{5/2}$ clock with $\Delta$ in a comfortable range~[see Fig.~\ref{fig:scheme}~(d)]. 
While both drive beams are derived from the same laser source, they are physically different at the location of the atoms, allowing ac-shift cancellation via separate intensity controls (as in Secs.~\ref{subsec:mode1} and~\ref{subsec:mode2}).

For this clock mode, it is $\Delta= 2 \pi \times 2.6 \times 10^3$~GHz. The ac polarizabilities are -16852 and 413 for $\vert 5S_{1/2}, * \rangle$ at $\lambda_{4,L}=774.985$~nm and $\lambda_{4,U}=1549.971$~nm, respectively, and -5 and -26080 for $\vert 4D_{5/2}, F''=4, m_{F''}=0 \rangle$ at the same wavelengths. These polarizabilities are from sums over all electric-dipole-coupled perturbing states, including $5P_{3/2}$. The polarizabilities yield 
a fixed Rabi-frequency ratio $\Omega_{PD}/\Omega_{SP}$ for which the clock-transition ac shift cancels.  
Requiring the same $\gamma_{4D} = 10^3$~s$^{-1}$ as in Secs.~\ref{subsec:mode1}-\ref{subsec:mode3}, Eq.~\ref{eq:g4D} then yields values $\Omega_{SP}= 2 \pi \times$~275~MHz and $\Omega_{PD}= 2 \pi \times$~319~MHz.
For beams with $w_0=1$~mm, the respective beam powers are 180~mW and 115~mW. These powers appear quite feasible for the required 775~nm/1550~nm wavelength combination. 
The individual clock-level ac shifts are both near $2 \pi \times$~8.92~kHz.  To achieve the limit of $2 \pi \times$ 60~Hz for the clock-transition shift, set in Sec.~\ref{sec:general}, the clock drive-beam intensities have to be controlled to within an imbalance of $\epsilon \approx 0.7 \%$, which is similar to the $\epsilon$-value assumed for the clock modes in Secs.~\ref{subsec:mode1} and~\ref{subsec:mode2}. 

Importantly, the 775~nm/1550~nm far-off-resonant $5S_{1/2}$ -$4D_{5/2}$ clock mode requires only a single laser source at 1550~nm; the 775-nm beam is generated with a frequency doubler. This leaves the 1550-nm laser as the only laser that must be tuned, which greatly simplifies the clock-laser architecture. Yet, with two beams of different colors being applied to the atoms, the method allows ac-shift cancellation. It also operates on a cycling transition regarding the relevant $F$ and $F''$-values. Further, both fluorescence 
wavelengths differ from both drive wavelengths by at least 5~nm, which suffices for high-contrast spectral filtering. The fluorescence 
wavelengths are both in spectral ranges for which excellent photodetectors exist (see Sec.~\ref{subsec:detectors}). 

Due to the advantages pointed out, the 775~nm/1550~nm far-off-resonant $5S_{1/2}$ - $4D_{5/2}$ clock mode is considered to be the most competitive among the four $4D_J$ clock modes discussed. 

\subsection{Far-off-resonant \texorpdfstring{$5S_{1/2}$ -$5D_{5/2}$}{} Doppler-free single-color two-photon drive}
\label{subsec:mode5}

We include a comparison with the ubiquitous Rb $5S_{1/2}$ - $5D_{5/2}$ Doppler-free two-photon clock. This clock has a more complex intermediate level scheme with 7 fine-structure states between the clock states. One typically measures the fluorescence cascade through the $6P_{3/2}$ level, which provides 420-nm fluorescence that can be filtered well from the infrared drive fields near 778~nm~\cite{martin2018, martinpra2019}. 
For the two-photon $5D_{5/2}$-clock, it is $\Delta= 2 \pi \times 1.06 \times 10^3$~GHz. Requiring $\gamma_{5D} = 10^3$~s$^{-1}$, in analogy with the value for $\gamma_{4D}$ in the previous sections, we find a laser power requirement of 161~mW for beams with $w_0=1$~mm, and the ac shift of the clock transition is -17.2~kHz, corresponding to a relative ac shift of the $5D_{5/2}$ clock's transition frequency of $-2.2 \times 10^{-13}/($mW/mm$^2$), which is close to the result of a rigorous calculation in~\cite{martinpra2019}. This ac shift is rather large and cannot be compensated in the Doppler-free one-color, two-photon configuration, making it one of the main drawbacks of the $5D_{5/2}$ clock. Additional notable disadvantages include a large black-body shift at 300~K of about $-2 \pi \times $~150~Hz, which is due to perturbing transitions in the 10-$\mu$m range, as well as second-order Zeeman and dc quadratic Stark shifts that are larger than for the $4D_J$ clocks (see Sec.~\ref{sec:syst}).

The lifetime of the $5D_{5/2}$ state exceeds that of the $4D_{5/2}$ state by about a factor of 2.6~\cite{safronova2011,shengpra2008,Heavens_61}, and the (total) clock frequency of the $5D_{5/2}$-clock is about a factor of 1.33 higher than that of the $4D_{5/2}$-clock. These facts amount to a clock-stability advantage for the $5D_{5/2}$-clock by a factor of 3.5, according to Eq.~\ref{eq:sigy} in the next section. However, the SNR for $5D_{5/2}$ decay is worse than for $4D_{5/2}$ decay because the 420-nm decay branch of $5D_{5/2}$ has a probability of only 30$\%$ and yields only one detectable photon (instead of two for $4D_{5/2}$-decay), and the quantum efficiency of photodetectors for 420-nm is only about half of that of detectors for $\sim$780~nm and $\sim$1500~nm. Those facts worsen the clock stability of the $5D_{5/2}$-clock by a factor of about $\sqrt{0.3 \times 0.5 \times 0.5} \approx 0.27$ relative to that of the
$4D_{5/2}$-clock. Hence, under the outlined assumptions the net stability advantage of the $5D_{5/2}$ over the $4D_{5/2}$ two-photon clock is about 0.9, {\sl{i.e.}}, the $4D_{5/2}$ clock would actually be marginally better. While this result is only an estimation, 
it stands to reason that stability disadvantages of the $4D_{5/2}$-clock due to larger linewidth and lower transition energy are compensated by advantages in the SNR. More details on clock stability are discussed in the next section.

We note that multi-color and relatively near-resonant implementations of Rb $5D_{5/2}$ clocks have been studied in~\cite{gerginov2018,perrella2019, hamilton2023}. Such implementations allow one to address clock-laser-induced ac shifts via differential intensity control.

\section{Statistical analysis}
\label{sec:statistics}

The Allan deviation of the relative clock frequency, commonly used to estimate the quantum-noise-limited clock stability, is often expressed as~\cite{atomicclocksreview, Tjoelker2016, Enzer2021}

\begin{equation}
    \sigma(\tau) = \frac{1}{\xi S} \frac{\Delta \nu_c}{\nu_c} \sqrt{\frac{T_m}{\tau}},
\label{eq:sigy}
\end{equation}

\noindent where $S$ is the SNR achieved in a single clock cycle, and $\Delta \nu_c$ is the full-width-at-half-maximum linewidth of the clock transition frequency in Hz. Further, $\nu_c$ is the frequency sum of the $5S_{1/2} \rightarrow 4D_J$ excitation lasers, $T_m$ is the measurement time for a single clock cycle, and $\tau$ is the total integration time. 
For clock lasers locked on a fringe of a Ramsey spectrum~\cite{Ramsey1950}, a case that is often considered, the linewidth $\Delta \nu_c$ is the full width at half maximum of the periodicity of the Ramsey spectrum in Hz. Quantum-projection noise~\cite{Itano1993} then yields an ideally lock-point-independent $\sigma(\tau)$ with $\xi=\pi$ in Eq.~\ref{eq:sigy}.  

In our case, $\Delta \nu_c$ is the inverse of the radiative lifetime of the upper clock state divided by $2 \pi$, equivalent to the low-saturation full width at half maximum of the clock transition line in Hz. The lifetime was recently determined to be 83~ns for $4D_{3/2}$ and 89~ns for $4D_{5/2}$, with less than 1~ns variation~\cite{safronova2004, Arora2012}. Hence, $\Delta \nu_c=1.92$~MHz and $\Delta \nu_c=1.78$~MHz for $4D_{3/2}$ and $4D_{5/2}$, respectively. The value of $\xi$ in Eq.~\ref{eq:sigy} varies depending on what model is adopted for the exact line shape. For a Gaussian with a peak quantum-state probability of 1 in the excited state, a somewhat un-physical case, we have found $\xi \approx 3.33$. For a mildly saturated Lorentzian that peaks at a quantum-state probability of 0.5 in the excited state, which is physically quite reasonable, we find $\xi \approx 1.41$. For any line shape model adopted, $\xi$  will depend on the detuning from the clock transition's line center, and it will typically become optimal at a detuning for which the excited-state probability is about one-half of its on-resonance peak value. Physical implementations of clock-laser locks will require a careful derivation of 
the factor $\xi$ in Eq.~\ref{eq:sigy}. For simplicity, in the following estimates we will use the commonly-quoted factor $\xi = \pi$ in Eq.~\ref{eq:sigy}. This means we assume that one can find a clock-laser lock scheme that performs as well as a quantum-projection-noise-limited lock to a fringe of the Ramsey spectrum of the clock transition.

For an estimate, we assume a flux 
$F_A= 10^7$~s$^{-1}$ of cold atoms passing through a clock probe region of 5~mm in length at a speed of 5~cm/s. The measurement time for a clock cycle equals the atom-field interaction time, $T_m = 0.1$~s. 
At the single-atom clock scattering rate of $\gamma_{4D} \sim 10^3$~s$^{-1}$ from Sec.~\ref{sec:scheme}, each atom provides 100 decays, the total rate of decays is $10^9$~s$^{-1}$, and the number of decays per $T_m$ is $10^8$. With an estimate of $\eta=10\%$ for the decay detection efficiency, the quantum-projection-limited SNR is $S=1/\sqrt{\eta N_P}=10^{7/2}$. From Eq.~\ref{eq:sigy} one then finds $\sigma(\tau) \approx 1.0 \times 10^{-13} / \sqrt{\tau}$. This is somewhat better than the demonstrated stability of the Rb $5D$ optical clock~\cite{martin2018}. 

It is noted that Eq.\ref{eq:sigy} in terms of the given practical parameters becomes 
\[\sigma(\tau) = \frac{1}{\xi \sqrt{\eta F_A \gamma_{4D} T_m \tau}} \frac{\Delta \nu_c}{\nu_c} \quad.\]
\noindent It is advantageous that the Rb~$4D_J$ fluorescence delivers two photons per decay at wavelengths for which Si and Ge photodiodes with near-peak efficiencies of $ \gtrsim 70 \%$ and with large areas exist. While still challenging, this will help achieving a decay detection efficiency of $\eta = 10\%$. 

\section{Detailed discussion of systematic shifts}
\label{sec:syst}

\subsection{Doppler effect}
\label{subsec:doppler}

With the exception of the single-color, two-photon Doppler-free clocks in Secs.~\ref{subsec:mode3} and~\ref{subsec:mode5},
the Doppler effect limits clock performance, a fact that has also been noted, to a lesser extent, in two-color $5D_{5/2}$-clocks~\cite{gerginov2018, perrella2019}. For counter-propagating excitation lasers with wavelengths as shown in Figs.~\ref{fig:scheme}~(a), (b) and (d), it is seen that the stability requirement of 60~Hz set in Sec.~\ref{sec:general} corresponds with an uncertainty of $\bar{v} \sim 0.1$~mm/s for the average velocity of the atom sample along the clock laser beam direction. 
At the same time, 
the velocity distribution can be $\sim 1$~m/s wide without substantially broadening the $4D_J$-clock lines, or about ten times the Doppler limit in Rb~\cite{Metcalf}. It is, however, challenging to laser-cool atom samples into velocity distributions with $\bar{v} \lesssim $~0.1~mm/s. For instance, radiation-pressure imbalance or magnetic fields in the laser-cooling region can cause $\bar{v} > 0.1$~mm/s. Also, radiation pressure from the clock lasers themselves must be avoided, as the recoil velocity for counter-propagating clock beams with wavelengths as in Secs.~\ref{subsec:mode2} and~\ref{subsec:mode4} is $\approx 3~$mm/s.

To solve this problem, here we consider a stream of cold atoms that moves along optical guiding channels. The guiding channels are about 1 MHz deep and are implemented by a two-dimensional (2D) optical lattice at a ``magic" wavelength (1060~nm; see Sec.~\ref{subsec:trappinglaser}). Atoms cooled to several tens of $\mu$K in a moving optical molasses~\cite{Metcalf} are adiabatically injected into the lattice channels, in which they travel at a mean forward speed of about 5~cm/s at a direction transverse to the lattice beams.
The clock interrogation region is defined by the overlap between the clock laser beams and the 2D-lattice channels. We envision clock-laser beams with $w_0$-waists in the range of $\sim 1$ to 5~mm, corresponding to probing times $T_m \lesssim 0.1$~s. To meet the condition $\vert {\bf{k}}_c \cdot \bar{{\bf{v}}} \vert \lesssim 0.1$~mm/s, with the clock wavevector ${\bf{k}}_c = {\bf{k}}_U - {\bf{k}}_L$ being the difference between upper- and lower-transition wavevectors, the 2D-lattice and the counter-propagating pairs of clock-laser beams are aligned in a plane with a precision of about 1~mrad. 

Along the ${\bf{k}}_c$-direction, the atoms are trapped in optical-lattice potential wells. For the aforementioned trap depth and wavelength, the center-of-mass (COM) oscillation frequency of the atoms in the wells is $f_{osc} \approx 100$~kHz,
or about 100 times the targeted clock scattering rate, $\gamma_{4D}$.  For the cases of Secs.~\ref{subsec:mode2} and~\ref{subsec:mode4} it is $k_c \approx 2 \pi/(1550$~nm$)$. In the harmonic approximation and in the Lamb-Dicke regime, $k_c x_0 << 1$ with $x_0 = \sqrt{h / (2 m f_{osc})}/(2 \pi)$, the in-trap clock spectrum consists of a Doppler-free carrier line and two motional side bands at frequency detunings $\pm f_{osc}$. The lower and upper side-band strengths relative to the carrier are $\approx (k_c x_0)^2 \, n$ and $\approx (k_c x_0)^2 \, (n+1)$, respectively, with COM quantum number $n$. The side bands are not resolved because $\Gamma_{4D} \gg f_{osc}$, and the line-pulling expression in Eq.~\ref{eq:pull} applies instead. One finds a fixed in-trap shift of the clock transition of $E_{rec} / \hbar \approx 2 \pi \times 955$~Hz, with the clock recoil energy $E_{rec} = (\hbar k_c)^2/(2 m)$. Also, the atom heating rate is $h \times 955$~Hz per clock excitation, or about $1\%$ of  $h \, f_{osc}$. We therefore believe that in-trap probing will effectively eliminate the Doppler effect in Rb $4D_J$ clocks.

Each atom is expected to undergo $\sim 100$ photon-scattering events during the clock-transition interrogation. As a precaution against radiation-pressure effects on the fluorescence from any un-trapped atoms, the clock beams and other relevant beams should be introduced in a radiation-pressure-neutral configuration. We envision sets of counter-propagating, intensity-matched pairs of beams for each color. This task will be eased by employing moderate-finesse linear optical cavities that provide both mode- and intensity-matched conditions, as in Fig.~\ref{fig:setup} below.

\subsection{Lattice-trapping laser}
\label{subsec:trappinglaser}

\begin{figure}[t!]
 \centering
  \includegraphics[width=0.48\textwidth]{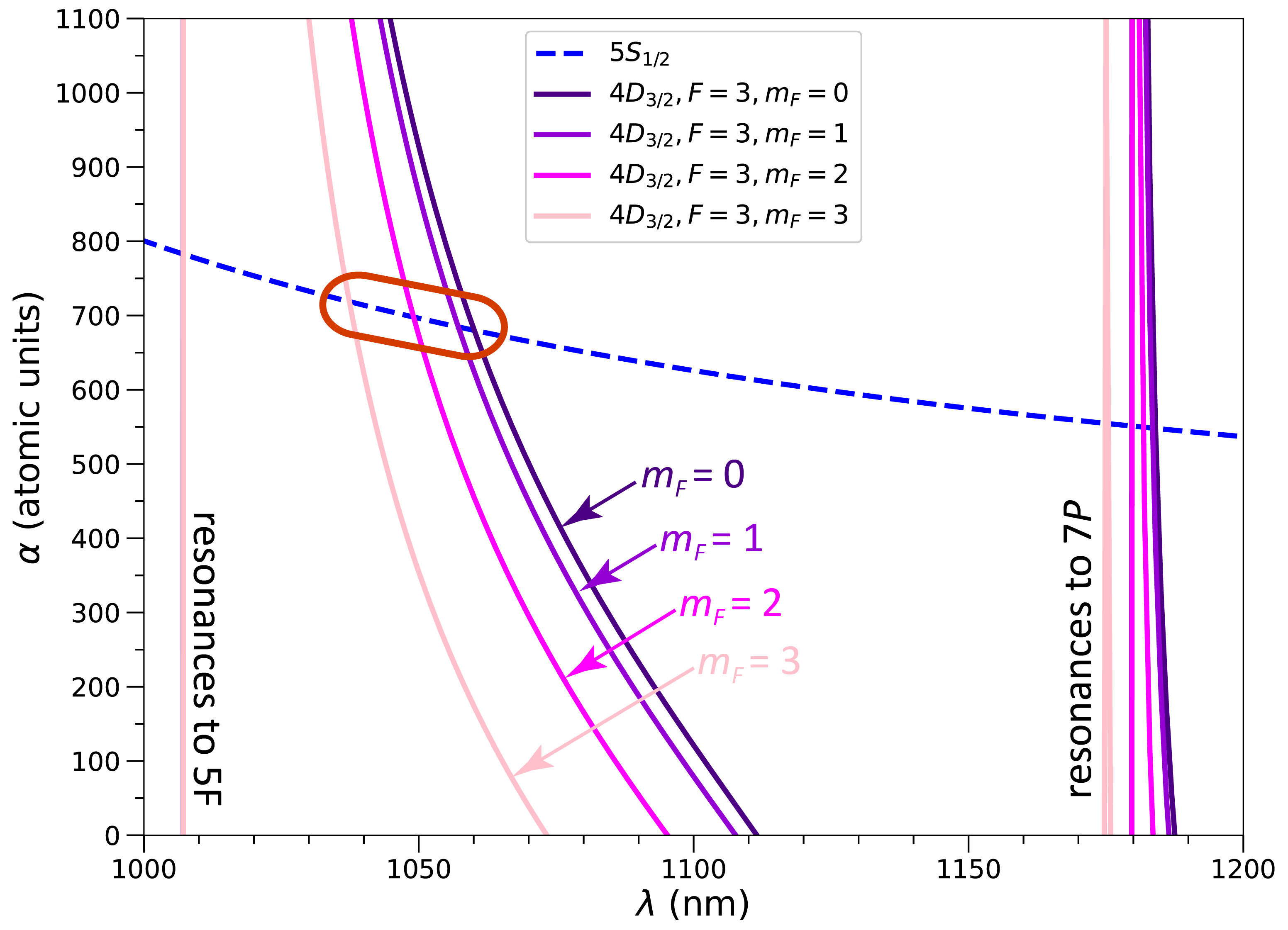}
  \caption{Calculated ac polarizabilities, $\alpha$, for the $\ket{5S_{1/2}}$ and $\ket{4D_{3/2}, F = 3, m_F}$ states in $^{87}$Rb. The region with ``magic" wavelength values is indicated. The resonances from the $4D_{3/2}$ state to the $5F$ and $7P$ states are indicated as well.} 
  \label{fig:pol}
\end{figure}

To probe the clock transition with the optical lattice left on, as assumed in Sec.~\ref{subsec:doppler}, 
the lower and upper clock states must have the same ac polarizability at the trapping wavelength to avoid clock-line shift due to the differential optical-lattice potential. To determine the ``magic" trapping wavelength, we obtain the ac polarizabilities, $\alpha$, using the same methods as in Sec.~\ref{sec:scheme}. The curves for $\alpha$ for the $5S_{1/2}$ and $4D_J$ states intersect within the region $\lambda~\in$~[1020, 1070]~nm, as shown in Fig.~\ref{fig:pol} for the case of the $4D_{3/2}$ state. At the level of precision considered, the ``magic" wavelengths and polarizabilities 
for $\vert 5S_{1/2}, * \rangle$, $\vert 4D_{3/2}, F=3, m_F=0 \rangle$,
and $\vert 4D_{5/2}, F=4, m_F=0 \rangle$ are $\lambda_{M} = 1060.1$~nm and $\alpha_{M} = 680$, respectively (polarizability in atomic units). At the ``magic" wavelength,
the polarizabilities for the next-higher $|m_F|$-states, $\vert 4D_{5/2}, F=4, m_F=1 \rangle$  and $\vert 4D_{3/2}, F=3, m_F=1 \rangle$,
differ from that for $m_F=0$ by 39 and 56, respectively, or about 6$\%$ and 8$\%$ of the lattice-induced shift. Since we estimate the accuracy of our polarizability values at $\lesssim 1\%$, the exact value of $\lambda_{M}$ may differ by $\lesssim 1$~nm from the value given. We expect that the exact value of $\lambda_{M}$ will have to be determined through precision measurement.

\begin{table*}[htb]
\begin{threeparttable}
\caption{\label{tab:table_shifts} Estimated systematic shifts for the proposed $4D_J$ clock schemes and a comparison with the $5D_{5/2}$ clock~\cite{martin2018,martinpra2019}}
\centering
\begin{tabular}{c c c c}
    \hline
    Source & $5S_{1/2} - 4D_{3/2}$ & $5S_{1/2} - 4D_{5/2}$ & $5S_{1/2} - 5D_{5/2}$ \\[1pt]
    \hline
    ac Stark shift due to excitation lasers~\tnote{a} & $<$60~Hz~\tnote{b} & $<$60~Hz/-15.7~kHz~\tnote{c} & -17.2~kHz \\[1pt]
    Second-order Zeeman shift~[kHz/G$^2$]~\tnote{d} & 8 & -24 & -53 \\[1pt]
    Doppler shift~[Hz] & 955~\tnote{e} & 955~\tnote{e} & 123~\tnote{f} \\[1pt]    
    BBR shift at 300~K~[Hz]~\tnote{g} & -2 & -2 & $<$~-150\\[1pt]
    dc Stark shift~[$\mu$Hz/(V/m)$^2$]~\tnote{h} & -4 & -3 & -228\\[1pt]
    \hline
\end{tabular}
     \begin{tablenotes}
       \item [a] Assuming a constant photon scattering rate $\gamma_{nD} = 10^3$~s$^{-1}$ and the respective laser-beam parameters discussed in Sec.~\ref{sec:scheme}
       \item [b] For both two-color excitation schemes in Sec.~\ref{subsec:mode1} and~\ref{subsec:mode2} 
       \item [c] Two / one-color excitation scheme in Sec.~\ref{subsec:mode4} / Sec.~\ref{subsec:mode3}    
       \item [d] Hyperfine constants of the respective excited states are from~\cite{moon2009,wang2014,Nez1993} 
       \item [e] Recoil clock shift in lattice, fixed
       \item [f] 2nd-order Doppler shift at 300~K        
       \item [g] Using our own calculations and information from~\cite{martinpra2019}
       \item [h] dc polarizabilities are from~\cite{UDportal}
     \end{tablenotes}
\end{threeparttable}
\end{table*}

For an optical lattice formed by counter-propagating beams with a peak trap depth of 1 MHz, from the ``magic" polarizability, $\alpha_{M}$=680, one finds a single-beam intensity of $I_1 = 78~$W/mm$^2$. 
Assuming that an optical resonator with a moderate finesse of $F \sim 300$ will be employed, for a Gaussian beam waist $w_0=$1~mm the laser power injected into the resonator would be $\lesssim 2$~W. Since at 
$\lambda_{M} \approx 1060$~nm high-power, narrow-band and tunable lasers are widely available, this power requirement appears reasonable. If necessary, one may increase $F$ or reduce $w_0$ to reduce the injected lattice power.   

Frequency fluctuations for the trap laser, $\Delta \nu_{trap}$, result in a variation of the differential polarizability between upper and lower clock states, and thus a variation of the clock transition frequency, $\Delta \nu_c$. For a full lattice depth of $V_0$ it is
\begin{equation}
\Delta \nu_c = \frac{V_0}{h c} \frac{\lambda_M^2}{\alpha_M} 
\Bigg|\frac{d \alpha_{4D}}{d\lambda} - \frac{d \alpha_{5S}}{d\lambda}\Bigg|_{\lambda_M} \Delta \nu_{trap} \quad.
\label{eq:traplaser}
\end{equation}
There, the derivatives of the clock-transition polarizabilities at $\lambda_M$ are -19.1/nm and -20.2/nm for the $4D_{3/2}$ and $4D_{5/2}$ clocks in Sec.~\ref{sec:scheme}. 
Requiring $| \Delta \nu_c | < 60~$Hz, the condition set in Sec. \ref{sec:general} to reach a $10^{-13}$ relative clock uncertainty, for a lattice depth of $V_0 = h \times 1$~MHz one finds from Eq.~\ref{eq:traplaser} a maximum allowed trap-laser frequency variation of about 500~MHz from the ``magic"-lattice condition. This number easily scales to other conditions, as it is inversely proportional to trap depth $V_0$ and proportional to the desired relative clock uncertainty.

\subsection{Second-order Zeeman shifts}
\label{subsec:bfield}

Next we consider the second-order Zeeman shifts in the bias field $B_{bias} \lesssim$100~mG, which is applied to maintain a well-defined quantization axis. The second-order Zeeman effects of the clock transition range from -52.6~kHz/G$^2$ for $5D_{5/2}$ (Sec.~\ref{subsec:mode5}) to -24.0~kHz/G$^2$ for $4D_{5/2}$ (Sec.~\ref{subsec:mode3} and~\ref{subsec:mode4}) and 7.7~kHz/G$^2$ for $4D_{3/2}$ (Sec.~\ref{subsec:mode1} and~\ref{subsec:mode2}), and are therefore comparatively benign. While the second-order Zeeman effect sets a slightly tighter limit for $5D$- than for $4D$-clocks due to the smaller hyperfine splittings of the $5D$-states, magnetic-field control at a level of about 10~mG, or $\lesssim 0.1 \, B_{bias}$ is sufficient to keep second-order Zeeman shifts below 60~Hz, the limit set in Sec.~\ref{sec:general}. 


\subsection{Black-body radiation}
\label{subsec:bbr}

Next, we consider clock-transition shifts induced by black-body radiation (BBR). Such shifts are important in a variety of optical atomic clocks~\cite{safronova_ieee_2012}, including a design based on the $5D$ state in Rb~\cite{martinpra2019}. In our BBR-shift estimate, we use methods from~\cite{farley1981} to find BBR shifts of Rb $5S_{1/2}$ and $4D_{3/2}$ at 300~K of $\approx$~-4.3 and $\approx$~-2.7~Hz, respectively, leading to a differential BBR shift on the clock transition of $\delta_{c, BBR}\approx$~-1.7~Hz. Thus, the clock discussed here should be robust against BBR effects and will not require additional infrastructure to compensate for them. For comparison, we obtain a differential shift for the $5S_{1/2} \rightarrow  5D_{5/2}$ transition of $\approx$~-155~Hz, in agreement with~\cite{martinpra2019}. The large BBR shift of $5D_{5/2}$ is due to a number of long-wavelength transitions into various $P$ and $F$ states (notably, $6P$ through $9P$ and $4F$ through $7F$), which overlap with the BBR spectrum at 300~K. The $4D_J$-states in Rb, in contrast, have no electric-dipole-allowed transitions at wavelengths longer than 2.3~$\mu$m.

\subsection{Stray dc electric fields and collisions}
\label{subsec:Edc}

For completeness, we estimate the fractional stability coefficients for the
quadratic dc Stark effect. The static polarizabilities from~\cite{UDportal} are $\approx$~6.8~$\times 10^{-17} /$~(V/cm)$^2$ 
and $\approx$~5.9~$\times 10^{-15} /$~(V/cm)$^2$ for $4D_{5/2}$ and $5D_{5/2}$ clocks, respectively, and the dc electric-field limits required for a fractional clock stability of $10^{-13}$ are 40~V/cm and 4~V/cm.  Hence, the dc Stark effect caused by stray electric fields is not expected to be a limiting factor for the low-lying states used in any of the proposed clock schemes. Stability limitations due to 
static electric fields may have to be assessed for clock implementations in miniature cells or vacuum systems, which may have significant contact or patch potentials.  

Shifts due to cold collisions in $^{87}$Rb are about a few Hz and, therefore, are not expected to be significant at the anticipated level of clock precision~\cite{gibble2000,sortais2000}.

\subsection{Summary of key systematics}
\label{subsec:table1}

In Table \ref{tab:table_shifts} we summarize several key systematics for Rb $4D$ and $5D$ clocks. For the Doppler shift, we list the in-trap photon recoil shift in two-color cold-atom $4D_J$ lattice clocks, as described in Secs.~\ref{subsec:mode1}, \ref{subsec:mode2} and~\ref{subsec:mode4}, and 
the second-order Doppler shift in Doppler-free $5D_{5/2}$ vapor-cell clocks from Sec.~\ref{subsec:mode5}. The in-trap photon recoil is a fixed-frequency offset of 955~Hz. The second-order Doppler shift in vapor cells is temperature-dependent.

\section{Discussion}
\label{sec:disc}

\begin{figure*}[htb]
 \centering
  \includegraphics[width=0.9\textwidth]{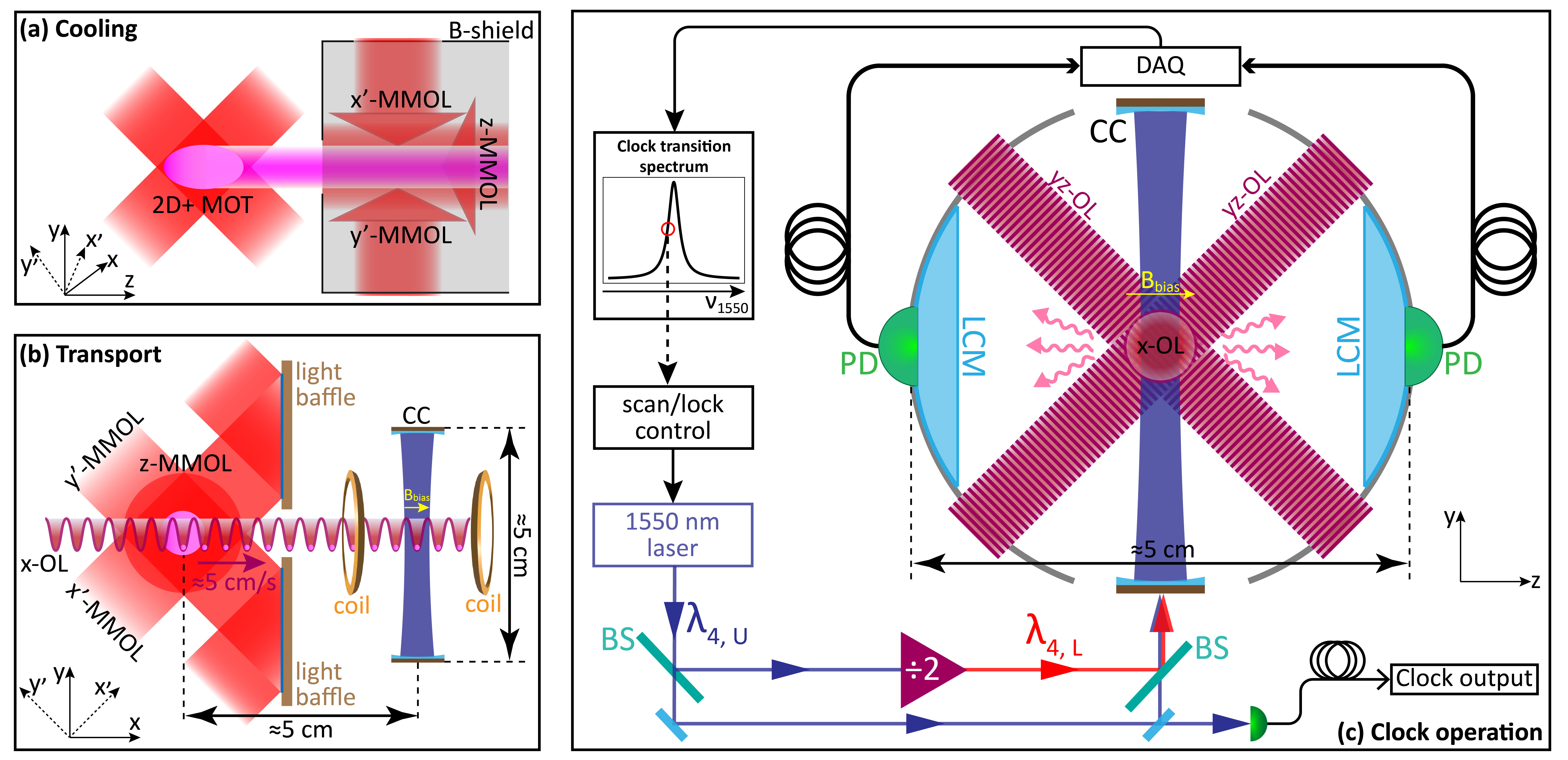}
  \caption{Architecture outline. (a) A cold atomic beam from a 2D+ MOT propagates along $z$ and passes through a magnetic shield into a moving optical molasses (MMOL). Four of the six molasses beams, along the $x'$- and $y'$-axes, have a frequency difference that promotes an atom flow along $x$ with an average speed of 0.1 m/s. (b) Atoms are trapped in a 1D 1060-nm conveyor optical lattice (x-OL) with a beam diameter of $\sim$~1 mm. The x-OL moves at a speed of 5~cm/s along $x$ through a light baffle. The bias field $B_{bias}$ points along $z$. (c) The x-OL intersects with the clock cavity (CC), which contains clock-drive laser beams of diameter $\sim 1$~mm, and 4 beams of a static 1060-nm optical lattice (yz-OL). Atoms on the x-OL conveyor are adiabatically introduced into the grid of OL-tubes, which have a spatial period of 530~nm, and are shuttled in the resultant moving 3D-lattice-trap through the CC, where the clock transition is interrogated. Clock fluorescence is collected through light-condensing mirrors (LCM) onto photodiodes (PD). The clock signal is used to lock the laser to the clock line. The two clock-laser beams are the fundamental and the second harmonic of a single narrow-band 1550-nm laser, which generates the fiber-coupled clock output. Optical pumping not shown.} 
  \label{fig:setup}
\end{figure*}

\subsection{Sample architecture}
\label{subsec:setup}

We finally outline a possible implementation of a Rb $4D_J$ clock in Fig.~\ref{fig:setup}, which is in-line with the estimates in Sec.~\ref{sec:statistics}. A 2D+~\cite{Dieckmann1998} or pyramidal~\cite{Camposeo2001} MOT supplies a cold atomic beam with a mean velocity of a few m/s along the $z$-direction and an average flux $\gtrsim 10^8$~s$^{-1}$. The atomic beam passes through a magnetic shield into a moving, red- or blue-detuned optical molasses~\cite{Metcalf}, which has a capture velocity sufficiently high to capture the majority of the cold atomic beam. Both sets of molasses beams in the $xy$-plane have identical frequency differences of about 100~kHz to maintain a flow of atoms along the $x$-direction (see Fig.~\ref{fig:setup}). The atoms are transferred into a $\sim 1$~MHz deep 1D optical lattice operating at the magic wavelength of about 1060.1~nm. The 1D lattice has a relative lattice-beam detuning such that the molasses and the 1D-lattice are co-moving at 5~cm/s along the $x$-direction, allowing for a seamless atom transfer into the moving 1D-lattice. 1D-lattice and molasses beams form an angle of $45^\circ$. The efficiency of the atom transfer is increased by dark-state extraction, where the re-pumper laser beam has a sharp drop-off realized by a knife-edge~\cite{Olson2006}. The atoms transferred into the 1D-lattice are shuttled out of the molasses region while being in the lower hyperfine state $F=1$, in which they do not scatter molasses light. Hence, the extraction proceeds without adverse radiation-pressure effects from the erratic fringe regions of the six moving-molasses laser beams. We may expect a flux of $F_A \gtrsim 10^7$~s$^{-1}$ atoms in the moving 1D-lattice at a temperature $\sim 10~\mu$K. The extracted atoms pass through a light baffle, which blocks molasses light from reaching the clock region.   

In the clock region, the atoms trapped in the moving 1D-lattice pass through the waist of a clock interrogation cavity. 
A set of four transverse optical-lattice beams operating near 1060.1~nm wavelength -- denoted yz-OL in Fig.~\ref{fig:setup} -- form a static 2D-lattice of atom guiding tubes. The lattice-trapped atoms propagate with a forward speed of 5~cm/s along these tubes through the clock probe region. The 2D-lattice beams and the clock-drive beams are carefully aligned in a plane with about 1~mrad tolerance for a Doppler-free clock drive, as discussed in Sec.~\ref{subsec:doppler}.

The lattice interrogation cavity allows for clock drive-field enhancement, mode cleanup, and radiation-pressure-neutral clock operation (see Fig.~\ref{fig:setup}). The clock cavity extends along the $y$-direction, and has a finesse of several 100 and a length of about 5~cm. For the beam waist we assume $w_0 \sim 5$~mm. A pair of Gaussian cavity modes at $\lambda_{4,L} \approx 775~$nm and $\lambda_{4,U} = 2\lambda_{4,L}$ drive the clock transition as described in Sec.~\ref{subsec:mode4}. The cavity has no resonant transverse modes that would degrade the intensity- and mode-matched profile of the counter-propagating clock drive fields applied to the atoms. The cavity is fine-aligned using in-vacuum piezo-electric actuators~\cite{Chen2014}, which allow one to tune cavity modes, which have $\sim 10$~MHz linewidth, into the $4D_{5/2}$ clock resonance. 
The clock may be operated at a reduced clock scattering rate $\gamma_{4D}$ without the clock cavity in place, using plain laser beams for the clock drive. 

The bias magnetic field, $B_{bias}$,  points along the $z$-axis and is applied by a pair of Helmholtz coils placed behind the magnetic shield. 

In the clock interrogation region, the probed atoms decay out of the $4D_{5/2}$ state with a rate of $\gamma_{4D}$ per atom. A pair of light-condensing mirrors concentrate the clock scattering light, which constitutes the clock signal to be measured, onto Si and Ge photodiodes. In advanced implementations, the mirrors are dichroic, with one mirror transmitting 780~nm and reflecting 1529~nm, and the other doing the opposite. The Si diodes are placed behind the 780-nm-transmitting condenser mirror, and the Ge diodes behind the 1529-nm-transmitting one. 
In this way, a maximum solid angle for bichromatic photon detection is achieved. The photodiodes are fitted with interference filters that block lattice, clock-drive, and other unwanted stray light. 

An important characteristic of the method in Sec.~\ref{subsec:mode4} is that the lower clock-drive beam is the second harmonic of the upper, as shown in Fig.~\ref{fig:setup}. In this way, a single laser operating near 1550~nm suffices to drive the clock. Among other advantages, in this scheme it is not necessary to stabilize two lasers 
in order to probe the $4D_{5//2}$ clock resonance. In addition to a greatly simplified overall clock-drive laser scheme, the single-laser design allows clock operation without the need for expensive testing equipment, such as high-finesse cavities or a frequency comb with phase-locked clock lasers. It is sufficient to lock the (only) 1550-nm clock laser to the Rb $4D_{5//2}$ clock resonance using a single laser lock. The evaluation of the stability and drift of the locked Rb $4D_{5/2}$ clock laser will then require an ultra-stable reference laser near $\lambda_{4,U}=1549.97$~nm, which is commercially available.   

\subsection{Conclusion}
\label{subsec:conclusion}

Considering clock stability according to 
Eq.~\ref{eq:sigy} and favorable systematics afforded by ac-shift cancellation, reduced black-body shifts, and reduced second-order dc-field shifts, we believe that the single-laser $4D_{5/2}$ 775~nm/1550~nm clock presents a good complement to the more widely employed $5D_{5/2}$ clock. 


In view of the fundamental physics properties described in our paper, Rb $4D_J$ clocks may serve well as stand-alone clocks in applications with moderate requirements (relative clock stability $\sim 10^{-13}/\sqrt{{\rm{Hz}}}$ and accuracy $\sim 10^{-13}$), or as a flywheel clock for ultra-high precision optical or nuclear clocks. All clock-excitation, laser-cooling, and ``magic"-lattice trapping lasers are readily available with the required power and laser linewidth specifications. Especially, the fundamental-color Rb $4D_J$ clock lasers are in the telecom S- and C-bands (1460 to 1530~nm). This fact could be exploited in long-distance clock linkage and quantum-networking applications. The clock-fluorescence photon yield of up to two photons per decaying atom, as well as the fluorescence colors, which are all in favored spectral ranges for which excellent photodetectors exist, are conducive to high clock bandwidth and SNR. Finally, with ongoing and rapid progress that is being made in low-SWaP and low-cost cold-atom techniques~(see, {\sl{e.g.}}, \cite{little2021, Strangfeld2021, mcgilligan2022, martinez2023, bregazzi2023, ropp2023}), we believe that the need for laser-cooled Rb atoms will become an increasingly less detrimental factor in future implementations of Rb $4D_J$ clocks. Components of the atom preparation, optical-lattice transfer, and 2D-lattice atom-guiding architecture presented in Sec.~\ref{subsec:setup} may be applicable to other ``magic"-lattice clocks, such as Sr and Yb clocks~\cite{atomicclocksreview}.

The discussed methods for high-precision spectroscopy of Rb $4D_J$ transitions at the 100-Hz level are also of interest in fundamental research on the properties of low-lying excited states. This includes hyperfine-coupling constants~\cite{moon2009, Duspayev2023}, lifetimes~\cite{Heavens_61}, and ac polarizabilities~\cite{cardman2021, atoms10040117} of the $4D_J$ states. The proposed Rb $4D_J$ optical-lattice clocks will require exact data on the ``magic" wavelengths of Rb $5S_{1/2}$ and $4D_J$ hyperfine states. High-precision measurements and ``magic" wavelengths will be of interest in comparisons with advanced atomic-structure calculations~\cite{safronova2011, trantan2023}. 

\maketitle
\section*{Acknowledgments}
\label{sec:acknowledgments}
We thank Dr. Ryan Cardman for useful discussions at the beginning of the study. This work was supported by the NSF Grant No. PHY-2110049. A.D. acknowledges support from the Rackham Predoctoral Fellowship at the University of Michigan.

\bibliography{references.bib}

\begin{thebibliography}{77}%
\makeatletter
\providecommand \@ifxundefined [1]{%
 \@ifx{#1\undefined}
}%
\providecommand \@ifnum [1]{%
 \ifnum #1\expandafter \@firstoftwo
 \else \expandafter \@secondoftwo
 \fi
}%
\providecommand \@ifx [1]{%
 \ifx #1\expandafter \@firstoftwo
 \else \expandafter \@secondoftwo
 \fi
}%
\providecommand \natexlab [1]{#1}%
\providecommand \enquote  [1]{``#1''}%
\providecommand \bibnamefont  [1]{#1}%
\providecommand \bibfnamefont [1]{#1}%
\providecommand \citenamefont [1]{#1}%
\providecommand \href@noop [0]{\@secondoftwo}%
\providecommand \href [0]{\begingroup \@sanitize@url \@href}%
\providecommand \@href[1]{\@@startlink{#1}\@@href}%
\providecommand \@@href[1]{\endgroup#1\@@endlink}%
\providecommand \@sanitize@url [0]{\catcode `\\12\catcode `\$12\catcode `\&12\catcode `\#12\catcode `\^12\catcode `\_12\catcode `\%12\relax}%
\providecommand \@@startlink[1]{}%
\providecommand \@@endlink[0]{}%
\providecommand \url  [0]{\begingroup\@sanitize@url \@url }%
\providecommand \@url [1]{\endgroup\@href {#1}{\urlprefix }}%
\providecommand \urlprefix  [0]{URL }%
\providecommand \Eprint [0]{\href }%
\providecommand \doibase [0]{https://doi.org/}%
\providecommand \selectlanguage [0]{\@gobble}%
\providecommand \bibinfo  [0]{\@secondoftwo}%
\providecommand \bibfield  [0]{\@secondoftwo}%
\providecommand \translation [1]{[#1]}%
\providecommand \BibitemOpen [0]{}%
\providecommand \bibitemStop [0]{}%
\providecommand \bibitemNoStop [0]{.\EOS\space}%
\providecommand \EOS [0]{\spacefactor3000\relax}%
\providecommand \BibitemShut  [1]{\csname bibitem#1\endcsname}%
\let\auto@bib@innerbib\@empty
\bibitem [{\citenamefont {Ludlow}\ \emph {et~al.}(2015)\citenamefont {Ludlow}, \citenamefont {Boyd}, \citenamefont {Ye}, \citenamefont {Peik},\ and\ \citenamefont {Schmidt}}]{atomicclocksreview}%
  \BibitemOpen
  \bibfield  {author} {\bibinfo {author} {\bibfnamefont {A.~D.}\ \bibnamefont {Ludlow}}, \bibinfo {author} {\bibfnamefont {M.~M.}\ \bibnamefont {Boyd}}, \bibinfo {author} {\bibfnamefont {J.}~\bibnamefont {Ye}}, \bibinfo {author} {\bibfnamefont {E.}~\bibnamefont {Peik}},\ and\ \bibinfo {author} {\bibfnamefont {P.~O.}\ \bibnamefont {Schmidt}},\ }\bibfield  {title} {\bibinfo {title} {Optical atomic clocks},\ }\href {https://doi.org/10.1103/RevModPhys.87.637} {\bibfield  {journal} {\bibinfo  {journal} {Rev. Mod. Phys.}\ }\textbf {\bibinfo {volume} {87}},\ \bibinfo {pages} {637} (\bibinfo {year} {2015})}\BibitemShut {NoStop}%
\bibitem [{\citenamefont {Kozlov}\ \emph {et~al.}(2018)\citenamefont {Kozlov}, \citenamefont {Safronova}, \citenamefont {Crespo L\'opez-Urrutia},\ and\ \citenamefont {Schmidt}}]{chargedclocksreview}%
  \BibitemOpen
  \bibfield  {author} {\bibinfo {author} {\bibfnamefont {M.~G.}\ \bibnamefont {Kozlov}}, \bibinfo {author} {\bibfnamefont {M.~S.}\ \bibnamefont {Safronova}}, \bibinfo {author} {\bibfnamefont {J.~R.}\ \bibnamefont {Crespo L\'opez-Urrutia}},\ and\ \bibinfo {author} {\bibfnamefont {P.~O.}\ \bibnamefont {Schmidt}},\ }\bibfield  {title} {\bibinfo {title} {Highly charged ions: Optical clocks and applications in fundamental physics},\ }\href {https://doi.org/10.1103/RevModPhys.90.045005} {\bibfield  {journal} {\bibinfo  {journal} {Rev. Mod. Phys.}\ }\textbf {\bibinfo {volume} {90}},\ \bibinfo {pages} {045005} (\bibinfo {year} {2018})}\BibitemShut {NoStop}%
\bibitem [{\citenamefont {Dimarcq}\ \emph {et~al.}(2024)\citenamefont {Dimarcq} \emph {et~al.}}]{Dimarcq_2024}%
  \BibitemOpen
  \bibfield  {author} {\bibinfo {author} {\bibfnamefont {N.}~\bibnamefont {Dimarcq}} \emph {et~al.},\ }\bibfield  {title} {\bibinfo {title} {Roadmap towards the redefinition of the second},\ }\href {https://doi.org/10.1088/1681-7575/ad17d2} {\bibfield  {journal} {\bibinfo  {journal} {Metrologia}\ }\textbf {\bibinfo {volume} {61}},\ \bibinfo {pages} {012001} (\bibinfo {year} {2024})}\BibitemShut {NoStop}%
\bibitem [{\citenamefont {Reinhardt}\ \emph {et~al.}(2007)\citenamefont {Reinhardt}, \citenamefont {Saathoff}, \citenamefont {Buhr}, \citenamefont {Carlson}, \citenamefont {Wolf}, \citenamefont {Schwalm}, \citenamefont {Karpuk}, \citenamefont {Novotny}, \citenamefont {Huber}, \citenamefont {Zimmermann}, \citenamefont {Holzwarth}, \citenamefont {Udem}, \citenamefont {Hänsch},\ and\ \citenamefont {Gwinner}}]{reinhardt2007}%
  \BibitemOpen
  \bibfield  {author} {\bibinfo {author} {\bibfnamefont {S.}~\bibnamefont {Reinhardt}}, \bibinfo {author} {\bibfnamefont {G.}~\bibnamefont {Saathoff}}, \bibinfo {author} {\bibfnamefont {H.}~\bibnamefont {Buhr}}, \bibinfo {author} {\bibfnamefont {L.~A.}\ \bibnamefont {Carlson}}, \bibinfo {author} {\bibfnamefont {A.}~\bibnamefont {Wolf}}, \bibinfo {author} {\bibfnamefont {D.}~\bibnamefont {Schwalm}}, \bibinfo {author} {\bibfnamefont {S.}~\bibnamefont {Karpuk}}, \bibinfo {author} {\bibfnamefont {C.}~\bibnamefont {Novotny}}, \bibinfo {author} {\bibfnamefont {G.}~\bibnamefont {Huber}}, \bibinfo {author} {\bibfnamefont {M.}~\bibnamefont {Zimmermann}}, \bibinfo {author} {\bibfnamefont {R.}~\bibnamefont {Holzwarth}}, \bibinfo {author} {\bibfnamefont {T.}~\bibnamefont {Udem}}, \bibinfo {author} {\bibfnamefont {T.~W.}\ \bibnamefont {Hänsch}},\ and\ \bibinfo {author} {\bibfnamefont {G.}~\bibnamefont {Gwinner}},\ }\bibfield  {title} {\bibinfo {title} {Test of relativistic time dilation with fast optical atomic clocks at
  different velocities},\ }\href {https://doi.org/10.1038/nphys778} {\bibfield  {journal} {\bibinfo  {journal} {Nat. Phys.}\ }\textbf {\bibinfo {volume} {3}},\ \bibinfo {pages} {861} (\bibinfo {year} {2007})}\BibitemShut {NoStop}%
\bibitem [{\citenamefont {Blatt}\ \emph {et~al.}(2008)\citenamefont {Blatt}, \citenamefont {Ludlow}, \citenamefont {Campbell}, \citenamefont {Thomsen}, \citenamefont {Zelevinsky}, \citenamefont {Boyd}, \citenamefont {Ye}, \citenamefont {Baillard}, \citenamefont {Fouch\'e}, \citenamefont {Le~Targat}, \citenamefont {Brusch}, \citenamefont {Lemonde}, \citenamefont {Takamoto}, \citenamefont {Hong}, \citenamefont {Katori},\ and\ \citenamefont {Flambaum}}]{blatt2008}%
  \BibitemOpen
  \bibfield  {author} {\bibinfo {author} {\bibfnamefont {S.}~\bibnamefont {Blatt}}, \bibinfo {author} {\bibfnamefont {A.~D.}\ \bibnamefont {Ludlow}}, \bibinfo {author} {\bibfnamefont {G.~K.}\ \bibnamefont {Campbell}}, \bibinfo {author} {\bibfnamefont {J.~W.}\ \bibnamefont {Thomsen}}, \bibinfo {author} {\bibfnamefont {T.}~\bibnamefont {Zelevinsky}}, \bibinfo {author} {\bibfnamefont {M.~M.}\ \bibnamefont {Boyd}}, \bibinfo {author} {\bibfnamefont {J.}~\bibnamefont {Ye}}, \bibinfo {author} {\bibfnamefont {X.}~\bibnamefont {Baillard}}, \bibinfo {author} {\bibfnamefont {M.}~\bibnamefont {Fouch\'e}}, \bibinfo {author} {\bibfnamefont {R.}~\bibnamefont {Le~Targat}}, \bibinfo {author} {\bibfnamefont {A.}~\bibnamefont {Brusch}}, \bibinfo {author} {\bibfnamefont {P.}~\bibnamefont {Lemonde}}, \bibinfo {author} {\bibfnamefont {M.}~\bibnamefont {Takamoto}}, \bibinfo {author} {\bibfnamefont {F.-L.}\ \bibnamefont {Hong}}, \bibinfo {author} {\bibfnamefont {H.}~\bibnamefont {Katori}},\ and\ \bibinfo {author} {\bibfnamefont
  {V.~V.}\ \bibnamefont {Flambaum}},\ }\bibfield  {title} {\bibinfo {title} {New limits on coupling of fundamental constants to gravity using $^{87}\mathrm{Sr}$ optical lattice clocks},\ }\href {https://doi.org/10.1103/PhysRevLett.100.140801} {\bibfield  {journal} {\bibinfo  {journal} {Phys. Rev. Lett.}\ }\textbf {\bibinfo {volume} {100}},\ \bibinfo {pages} {140801} (\bibinfo {year} {2008})}\BibitemShut {NoStop}%
\bibitem [{\citenamefont {Kolkowitz}\ \emph {et~al.}(2016)\citenamefont {Kolkowitz}, \citenamefont {Pikovski}, \citenamefont {Langellier}, \citenamefont {Lukin}, \citenamefont {Walsworth},\ and\ \citenamefont {Ye}}]{kolkowitz2016}%
  \BibitemOpen
  \bibfield  {author} {\bibinfo {author} {\bibfnamefont {S.}~\bibnamefont {Kolkowitz}}, \bibinfo {author} {\bibfnamefont {I.}~\bibnamefont {Pikovski}}, \bibinfo {author} {\bibfnamefont {N.}~\bibnamefont {Langellier}}, \bibinfo {author} {\bibfnamefont {M.~D.}\ \bibnamefont {Lukin}}, \bibinfo {author} {\bibfnamefont {R.~L.}\ \bibnamefont {Walsworth}},\ and\ \bibinfo {author} {\bibfnamefont {J.}~\bibnamefont {Ye}},\ }\bibfield  {title} {\bibinfo {title} {Gravitational wave detection with optical lattice atomic clocks},\ }\href {https://doi.org/10.1103/PhysRevD.94.124043} {\bibfield  {journal} {\bibinfo  {journal} {Phys. Rev. D}\ }\textbf {\bibinfo {volume} {94}},\ \bibinfo {pages} {124043} (\bibinfo {year} {2016})}\BibitemShut {NoStop}%
\bibitem [{\citenamefont {Derevianko}\ and\ \citenamefont {Pospelov}(2014)}]{derevianko2014}%
  \BibitemOpen
  \bibfield  {author} {\bibinfo {author} {\bibfnamefont {A.}~\bibnamefont {Derevianko}}\ and\ \bibinfo {author} {\bibfnamefont {M.}~\bibnamefont {Pospelov}},\ }\bibfield  {title} {\bibinfo {title} {Hunting for topological dark matter with atomic clocks},\ }\href {https://doi.org/10.1038/nphys3137} {\bibfield  {journal} {\bibinfo  {journal} {Nat. Phys.}\ }\textbf {\bibinfo {volume} {10}},\ \bibinfo {pages} {933} (\bibinfo {year} {2014})}\BibitemShut {NoStop}%
\bibitem [{\citenamefont {Wcis{\l}o}\ \emph {et~al.}(2018)\citenamefont {Wcis{\l}o}, \citenamefont {Ablewski}, \citenamefont {Beloy}, \citenamefont {Bilicki}, \citenamefont {Bober}, \citenamefont {Brown}, \citenamefont {Fasano}, \citenamefont {Ciury{\l}o}, \citenamefont {Hachisu}, \citenamefont {Ido} \emph {et~al.}}]{wcislo2018}%
  \BibitemOpen
  \bibfield  {author} {\bibinfo {author} {\bibfnamefont {P.}~\bibnamefont {Wcis{\l}o}}, \bibinfo {author} {\bibfnamefont {P.}~\bibnamefont {Ablewski}}, \bibinfo {author} {\bibfnamefont {K.}~\bibnamefont {Beloy}}, \bibinfo {author} {\bibfnamefont {S.}~\bibnamefont {Bilicki}}, \bibinfo {author} {\bibfnamefont {M.}~\bibnamefont {Bober}}, \bibinfo {author} {\bibfnamefont {R.}~\bibnamefont {Brown}}, \bibinfo {author} {\bibfnamefont {R.}~\bibnamefont {Fasano}}, \bibinfo {author} {\bibfnamefont {R.}~\bibnamefont {Ciury{\l}o}}, \bibinfo {author} {\bibfnamefont {H.}~\bibnamefont {Hachisu}}, \bibinfo {author} {\bibfnamefont {T.}~\bibnamefont {Ido}}, \emph {et~al.},\ }\bibfield  {title} {\bibinfo {title} {New bounds on dark matter coupling from a global network of optical atomic clocks},\ }\href {https://doi.org/10.1126/sciadv.aau4869} {\bibfield  {journal} {\bibinfo  {journal} {Sci. Adv.}\ }\textbf {\bibinfo {volume} {4}},\ \bibinfo {pages} {eaau4869} (\bibinfo {year} {2018})}\BibitemShut {NoStop}%
\bibitem [{\citenamefont {Kobayashi}\ \emph {et~al.}(2022)\citenamefont {Kobayashi}, \citenamefont {Takamizawa}, \citenamefont {Akamatsu}, \citenamefont {Kawasaki}, \citenamefont {Nishiyama}, \citenamefont {Hosaka}, \citenamefont {Hisai}, \citenamefont {Wada}, \citenamefont {Inaba}, \citenamefont {Tanabe},\ and\ \citenamefont {Yasuda}}]{kobayashi2022}%
  \BibitemOpen
  \bibfield  {author} {\bibinfo {author} {\bibfnamefont {T.}~\bibnamefont {Kobayashi}}, \bibinfo {author} {\bibfnamefont {A.}~\bibnamefont {Takamizawa}}, \bibinfo {author} {\bibfnamefont {D.}~\bibnamefont {Akamatsu}}, \bibinfo {author} {\bibfnamefont {A.}~\bibnamefont {Kawasaki}}, \bibinfo {author} {\bibfnamefont {A.}~\bibnamefont {Nishiyama}}, \bibinfo {author} {\bibfnamefont {K.}~\bibnamefont {Hosaka}}, \bibinfo {author} {\bibfnamefont {Y.}~\bibnamefont {Hisai}}, \bibinfo {author} {\bibfnamefont {M.}~\bibnamefont {Wada}}, \bibinfo {author} {\bibfnamefont {H.}~\bibnamefont {Inaba}}, \bibinfo {author} {\bibfnamefont {T.}~\bibnamefont {Tanabe}},\ and\ \bibinfo {author} {\bibfnamefont {M.}~\bibnamefont {Yasuda}},\ }\bibfield  {title} {\bibinfo {title} {Search for ultralight dark matter from long-term frequency comparisons of optical and microwave atomic clocks},\ }\href {https://doi.org/10.1103/PhysRevLett.129.241301} {\bibfield  {journal} {\bibinfo  {journal} {Phys. Rev. Lett.}\ }\textbf {\bibinfo {volume}
  {129}},\ \bibinfo {pages} {241301} (\bibinfo {year} {2022})}\BibitemShut {NoStop}%
\bibitem [{\citenamefont {Filzinger}\ \emph {et~al.}(2023)\citenamefont {Filzinger}, \citenamefont {D\"orscher}, \citenamefont {Lange}, \citenamefont {Klose}, \citenamefont {Steinel}, \citenamefont {Benkler}, \citenamefont {Peik}, \citenamefont {Lisdat},\ and\ \citenamefont {Huntemann}}]{filzinger2023}%
  \BibitemOpen
  \bibfield  {author} {\bibinfo {author} {\bibfnamefont {M.}~\bibnamefont {Filzinger}}, \bibinfo {author} {\bibfnamefont {S.}~\bibnamefont {D\"orscher}}, \bibinfo {author} {\bibfnamefont {R.}~\bibnamefont {Lange}}, \bibinfo {author} {\bibfnamefont {J.}~\bibnamefont {Klose}}, \bibinfo {author} {\bibfnamefont {M.}~\bibnamefont {Steinel}}, \bibinfo {author} {\bibfnamefont {E.}~\bibnamefont {Benkler}}, \bibinfo {author} {\bibfnamefont {E.}~\bibnamefont {Peik}}, \bibinfo {author} {\bibfnamefont {C.}~\bibnamefont {Lisdat}},\ and\ \bibinfo {author} {\bibfnamefont {N.}~\bibnamefont {Huntemann}},\ }\bibfield  {title} {\bibinfo {title} {Improved limits on the coupling of ultralight bosonic dark matter to photons from optical atomic clock comparisons},\ }\href {https://doi.org/10.1103/PhysRevLett.130.253001} {\bibfield  {journal} {\bibinfo  {journal} {Phys. Rev. Lett.}\ }\textbf {\bibinfo {volume} {130}},\ \bibinfo {pages} {253001} (\bibinfo {year} {2023})}\BibitemShut {NoStop}%
\bibitem [{\citenamefont {Leschiutta}(2005)}]{Leschiutta_2005}%
  \BibitemOpen
  \bibfield  {author} {\bibinfo {author} {\bibfnamefont {S.}~\bibnamefont {Leschiutta}},\ }\bibfield  {title} {\bibinfo {title} {The definition of the ‘atomic’ second},\ }\href {https://doi.org/10.1088/0026-1394/42/3/S03} {\bibfield  {journal} {\bibinfo  {journal} {Metrologia}\ }\textbf {\bibinfo {volume} {42}},\ \bibinfo {pages} {S10} (\bibinfo {year} {2005})}\BibitemShut {NoStop}%
\bibitem [{\citenamefont {Heavner}\ \emph {et~al.}(2014)\citenamefont {Heavner}, \citenamefont {Donley}, \citenamefont {Levi}, \citenamefont {Costanzo}, \citenamefont {Parker}, \citenamefont {Shirley}, \citenamefont {Ashby}, \citenamefont {Barlow},\ and\ \citenamefont {Jefferts}}]{Heavner2014}%
  \BibitemOpen
  \bibfield  {author} {\bibinfo {author} {\bibfnamefont {T.~P.}\ \bibnamefont {Heavner}}, \bibinfo {author} {\bibfnamefont {E.~A.}\ \bibnamefont {Donley}}, \bibinfo {author} {\bibfnamefont {F.}~\bibnamefont {Levi}}, \bibinfo {author} {\bibfnamefont {G.}~\bibnamefont {Costanzo}}, \bibinfo {author} {\bibfnamefont {T.~E.}\ \bibnamefont {Parker}}, \bibinfo {author} {\bibfnamefont {J.~H.}\ \bibnamefont {Shirley}}, \bibinfo {author} {\bibfnamefont {N.}~\bibnamefont {Ashby}}, \bibinfo {author} {\bibfnamefont {S.}~\bibnamefont {Barlow}},\ and\ \bibinfo {author} {\bibfnamefont {S.~R.}\ \bibnamefont {Jefferts}},\ }\bibfield  {title} {\bibinfo {title} {First accuracy evaluation of {NIST-F2}},\ }\href {https://doi.org/10.1088/0026-1394/51/3/174} {\bibfield  {journal} {\bibinfo  {journal} {Metrologia}\ }\textbf {\bibinfo {volume} {51}},\ \bibinfo {pages} {174} (\bibinfo {year} {2014})}\BibitemShut {NoStop}%
\bibitem [{\citenamefont {Beattie}\ \emph {et~al.}(2020)\citenamefont {Beattie}, \citenamefont {Jian}, \citenamefont {Alcock}, \citenamefont {Gertsvolf}, \citenamefont {Hendricks}, \citenamefont {Szymaniec},\ and\ \citenamefont {Gibble}}]{Beattie2020}%
  \BibitemOpen
  \bibfield  {author} {\bibinfo {author} {\bibfnamefont {S.}~\bibnamefont {Beattie}}, \bibinfo {author} {\bibfnamefont {B.}~\bibnamefont {Jian}}, \bibinfo {author} {\bibfnamefont {J.}~\bibnamefont {Alcock}}, \bibinfo {author} {\bibfnamefont {M.}~\bibnamefont {Gertsvolf}}, \bibinfo {author} {\bibfnamefont {R.}~\bibnamefont {Hendricks}}, \bibinfo {author} {\bibfnamefont {K.}~\bibnamefont {Szymaniec}},\ and\ \bibinfo {author} {\bibfnamefont {K.}~\bibnamefont {Gibble}},\ }\bibfield  {title} {\bibinfo {title} {First accuracy evaluation of the {NRC-FCs2} primary frequency standard},\ }\href {https://doi.org/10.1088/1681-7575/ab7c54} {\bibfield  {journal} {\bibinfo  {journal} {Metrologia}\ }\textbf {\bibinfo {volume} {57}},\ \bibinfo {pages} {035010} (\bibinfo {year} {2020})}\BibitemShut {NoStop}%
\bibitem [{\citenamefont {Kim}\ \emph {et~al.}(2023)\citenamefont {Kim}, \citenamefont {Aeppli}, \citenamefont {Bothwell},\ and\ \citenamefont {Ye}}]{kimprl2023}%
  \BibitemOpen
  \bibfield  {author} {\bibinfo {author} {\bibfnamefont {K.}~\bibnamefont {Kim}}, \bibinfo {author} {\bibfnamefont {A.}~\bibnamefont {Aeppli}}, \bibinfo {author} {\bibfnamefont {T.}~\bibnamefont {Bothwell}},\ and\ \bibinfo {author} {\bibfnamefont {J.}~\bibnamefont {Ye}},\ }\bibfield  {title} {\bibinfo {title} {Evaluation of lattice light shift at low ${10}^{\ensuremath{-}19}$ uncertainty for a shallow lattice {S}r optical clock},\ }\href {https://doi.org/10.1103/PhysRevLett.130.113203} {\bibfield  {journal} {\bibinfo  {journal} {Phys. Rev. Lett.}\ }\textbf {\bibinfo {volume} {130}},\ \bibinfo {pages} {113203} (\bibinfo {year} {2023})}\BibitemShut {NoStop}%
\bibitem [{\citenamefont {Bothwell}\ \emph {et~al.}(2022)\citenamefont {Bothwell}, \citenamefont {Kennedy}, \citenamefont {Aeppli}, \citenamefont {Kedar}, \citenamefont {Robinson}, \citenamefont {Oelker}, \citenamefont {Staron},\ and\ \citenamefont {Ye}}]{bothwell2022}%
  \BibitemOpen
  \bibfield  {author} {\bibinfo {author} {\bibfnamefont {T.}~\bibnamefont {Bothwell}}, \bibinfo {author} {\bibfnamefont {C.~J.}\ \bibnamefont {Kennedy}}, \bibinfo {author} {\bibfnamefont {A.}~\bibnamefont {Aeppli}}, \bibinfo {author} {\bibfnamefont {D.}~\bibnamefont {Kedar}}, \bibinfo {author} {\bibfnamefont {J.~M.}\ \bibnamefont {Robinson}}, \bibinfo {author} {\bibfnamefont {E.}~\bibnamefont {Oelker}}, \bibinfo {author} {\bibfnamefont {A.}~\bibnamefont {Staron}},\ and\ \bibinfo {author} {\bibfnamefont {J.}~\bibnamefont {Ye}},\ }\bibfield  {title} {\bibinfo {title} {Resolving the gravitational redshift across a millimetre-scale atomic sample},\ }\href {https://doi.org/10.1038/s41586-021-04349-7} {\bibfield  {journal} {\bibinfo  {journal} {Nature}\ }\textbf {\bibinfo {volume} {602}},\ \bibinfo {pages} {420} (\bibinfo {year} {2022})}\BibitemShut {NoStop}%
\bibitem [{\citenamefont {Zheng}\ \emph {et~al.}(2022)\citenamefont {Zheng}, \citenamefont {Dolde}, \citenamefont {Lochab}, \citenamefont {Merriman}, \citenamefont {Li},\ and\ \citenamefont {Kolkowitz}}]{zheng2022}%
  \BibitemOpen
  \bibfield  {author} {\bibinfo {author} {\bibfnamefont {X.}~\bibnamefont {Zheng}}, \bibinfo {author} {\bibfnamefont {J.}~\bibnamefont {Dolde}}, \bibinfo {author} {\bibfnamefont {V.}~\bibnamefont {Lochab}}, \bibinfo {author} {\bibfnamefont {B.~N.}\ \bibnamefont {Merriman}}, \bibinfo {author} {\bibfnamefont {H.}~\bibnamefont {Li}},\ and\ \bibinfo {author} {\bibfnamefont {S.}~\bibnamefont {Kolkowitz}},\ }\bibfield  {title} {\bibinfo {title} {Differential clock comparisons with a multiplexed optical lattice clock},\ }\href {https://doi.org/10.1038/s41586-021-04344-y} {\bibfield  {journal} {\bibinfo  {journal} {Nature}\ }\textbf {\bibinfo {volume} {602}},\ \bibinfo {pages} {425} (\bibinfo {year} {2022})}\BibitemShut {NoStop}%
\bibitem [{\citenamefont {Huntemann}\ \emph {et~al.}(2016)\citenamefont {Huntemann}, \citenamefont {Sanner}, \citenamefont {Lipphardt}, \citenamefont {Tamm},\ and\ \citenamefont {Peik}}]{huntemann2016}%
  \BibitemOpen
  \bibfield  {author} {\bibinfo {author} {\bibfnamefont {N.}~\bibnamefont {Huntemann}}, \bibinfo {author} {\bibfnamefont {C.}~\bibnamefont {Sanner}}, \bibinfo {author} {\bibfnamefont {B.}~\bibnamefont {Lipphardt}}, \bibinfo {author} {\bibfnamefont {C.}~\bibnamefont {Tamm}},\ and\ \bibinfo {author} {\bibfnamefont {E.}~\bibnamefont {Peik}},\ }\bibfield  {title} {\bibinfo {title} {Single-ion atomic clock with $3\ifmmode\times\else\texttimes\fi{}{10}^{\ensuremath{-}18}$ systematic uncertainty},\ }\href {https://doi.org/10.1103/PhysRevLett.116.063001} {\bibfield  {journal} {\bibinfo  {journal} {Phys. Rev. Lett.}\ }\textbf {\bibinfo {volume} {116}},\ \bibinfo {pages} {063001} (\bibinfo {year} {2016})}\BibitemShut {NoStop}%
\bibitem [{\citenamefont {McGrew}\ \emph {et~al.}(2018)\citenamefont {McGrew}, \citenamefont {Zhang}, \citenamefont {Fasano}, \citenamefont {Schäffer}, \citenamefont {Beloy}, \citenamefont {Nicolodi}, \citenamefont {Brown}, \citenamefont {Hinkley}, \citenamefont {Milani}, \citenamefont {Schioppo}, \citenamefont {Yoon},\ and\ \citenamefont {Ludlow}}]{mcgrew2018}%
  \BibitemOpen
  \bibfield  {author} {\bibinfo {author} {\bibfnamefont {W.~F.}\ \bibnamefont {McGrew}}, \bibinfo {author} {\bibfnamefont {X.}~\bibnamefont {Zhang}}, \bibinfo {author} {\bibfnamefont {R.~J.}\ \bibnamefont {Fasano}}, \bibinfo {author} {\bibfnamefont {S.~A.}\ \bibnamefont {Schäffer}}, \bibinfo {author} {\bibfnamefont {K.}~\bibnamefont {Beloy}}, \bibinfo {author} {\bibfnamefont {D.}~\bibnamefont {Nicolodi}}, \bibinfo {author} {\bibfnamefont {R.~C.}\ \bibnamefont {Brown}}, \bibinfo {author} {\bibfnamefont {N.}~\bibnamefont {Hinkley}}, \bibinfo {author} {\bibfnamefont {G.}~\bibnamefont {Milani}}, \bibinfo {author} {\bibfnamefont {M.}~\bibnamefont {Schioppo}}, \bibinfo {author} {\bibfnamefont {T.~H.}\ \bibnamefont {Yoon}},\ and\ \bibinfo {author} {\bibfnamefont {A.~D.}\ \bibnamefont {Ludlow}},\ }\bibfield  {title} {\bibinfo {title} {Atomic clock performance enabling geodesy below the centimetre level},\ }\href {https://doi.org/10.1038/s41586-018-0738-2} {\bibfield  {journal} {\bibinfo  {journal} {Nature}\ }\textbf
  {\bibinfo {volume} {564}},\ \bibinfo {pages} {87} (\bibinfo {year} {2018})}\BibitemShut {NoStop}%
\bibitem [{\citenamefont {Seiferle}\ \emph {et~al.}(2018)\citenamefont {Seiferle}, \citenamefont {von~der Wense}, \citenamefont {Bilous}, \citenamefont {Amersdorffer}, \citenamefont {Lemell}, \citenamefont {Libisch}, \citenamefont {Stellmer}, \citenamefont {Schumm}, \citenamefont {Düllmann}, \citenamefont {Pálffy},\ and\ \citenamefont {Thirolf}}]{seiferle2018}%
  \BibitemOpen
  \bibfield  {author} {\bibinfo {author} {\bibfnamefont {B.}~\bibnamefont {Seiferle}}, \bibinfo {author} {\bibfnamefont {L.}~\bibnamefont {von~der Wense}}, \bibinfo {author} {\bibfnamefont {P.~V.}\ \bibnamefont {Bilous}}, \bibinfo {author} {\bibfnamefont {I.}~\bibnamefont {Amersdorffer}}, \bibinfo {author} {\bibfnamefont {C.}~\bibnamefont {Lemell}}, \bibinfo {author} {\bibfnamefont {F.}~\bibnamefont {Libisch}}, \bibinfo {author} {\bibfnamefont {S.}~\bibnamefont {Stellmer}}, \bibinfo {author} {\bibfnamefont {T.}~\bibnamefont {Schumm}}, \bibinfo {author} {\bibfnamefont {C.~E.}\ \bibnamefont {Düllmann}}, \bibinfo {author} {\bibfnamefont {A.}~\bibnamefont {Pálffy}},\ and\ \bibinfo {author} {\bibfnamefont {P.~G.}\ \bibnamefont {Thirolf}},\ }\bibfield  {title} {\bibinfo {title} {Energy of the 229{T}h nuclear clock transition},\ }\href {https://doi.org/10.1038/s41586-019-1533-4} {\bibfield  {journal} {\bibinfo  {journal} {Nature}\ }\textbf {\bibinfo {volume} {573}},\ \bibinfo {pages} {243} (\bibinfo {year}
  {2018})}\BibitemShut {NoStop}%
\bibitem [{\citenamefont {Holliman}\ \emph {et~al.}(2022)\citenamefont {Holliman}, \citenamefont {Fan}, \citenamefont {Contractor}, \citenamefont {Brewer},\ and\ \citenamefont {Jayich}}]{holliman2022}%
  \BibitemOpen
  \bibfield  {author} {\bibinfo {author} {\bibfnamefont {C.~A.}\ \bibnamefont {Holliman}}, \bibinfo {author} {\bibfnamefont {M.}~\bibnamefont {Fan}}, \bibinfo {author} {\bibfnamefont {A.}~\bibnamefont {Contractor}}, \bibinfo {author} {\bibfnamefont {S.~M.}\ \bibnamefont {Brewer}},\ and\ \bibinfo {author} {\bibfnamefont {A.~M.}\ \bibnamefont {Jayich}},\ }\bibfield  {title} {\bibinfo {title} {Radium ion optical clock},\ }\href {https://doi.org/10.1103/PhysRevLett.128.033202} {\bibfield  {journal} {\bibinfo  {journal} {Phys. Rev. Lett.}\ }\textbf {\bibinfo {volume} {128}},\ \bibinfo {pages} {033202} (\bibinfo {year} {2022})}\BibitemShut {NoStop}%
\bibitem [{\citenamefont {Zhiqiang}\ \emph {et~al.}(2023)\citenamefont {Zhiqiang}, \citenamefont {Arnold}, \citenamefont {Kaewuam},\ and\ \citenamefont {Barrett}}]{zhiqiang2023}%
  \BibitemOpen
  \bibfield  {author} {\bibinfo {author} {\bibfnamefont {Z.}~\bibnamefont {Zhiqiang}}, \bibinfo {author} {\bibfnamefont {K.~J.}\ \bibnamefont {Arnold}}, \bibinfo {author} {\bibfnamefont {R.}~\bibnamefont {Kaewuam}},\ and\ \bibinfo {author} {\bibfnamefont {M.~D.}\ \bibnamefont {Barrett}},\ }\bibfield  {title} {\bibinfo {title} {$^{176}${Lu} clock comparison at the 10$^{-18}$ level via correlation spectroscopy},\ }\href {https://doi.org/10.1126/sciadv.adg1971} {\bibfield  {journal} {\bibinfo  {journal} {Sci. Adv.}\ }\textbf {\bibinfo {volume} {9}},\ \bibinfo {pages} {eadg1971} (\bibinfo {year} {2023})}\BibitemShut {NoStop}%
\bibitem [{\citenamefont {Mehlstäubler}\ \emph {et~al.}(2018)\citenamefont {Mehlstäubler}, \citenamefont {Grosche}, \citenamefont {Lisdat}, \citenamefont {Schmidt},\ and\ \citenamefont {Denker}}]{Mehlstäubler_2018}%
  \BibitemOpen
  \bibfield  {author} {\bibinfo {author} {\bibfnamefont {T.~E.}\ \bibnamefont {Mehlstäubler}}, \bibinfo {author} {\bibfnamefont {G.}~\bibnamefont {Grosche}}, \bibinfo {author} {\bibfnamefont {C.}~\bibnamefont {Lisdat}}, \bibinfo {author} {\bibfnamefont {P.~O.}\ \bibnamefont {Schmidt}},\ and\ \bibinfo {author} {\bibfnamefont {H.}~\bibnamefont {Denker}},\ }\bibfield  {title} {\bibinfo {title} {Atomic clocks for geodesy},\ }\href {https://doi.org/10.1088/1361-6633/aab409} {\bibfield  {journal} {\bibinfo  {journal} {Rep. Prog. Phys.}\ }\textbf {\bibinfo {volume} {81}},\ \bibinfo {pages} {064401} (\bibinfo {year} {2018})}\BibitemShut {NoStop}%
\bibitem [{\citenamefont {Ely}\ \emph {et~al.}(2018)\citenamefont {Ely}, \citenamefont {Burt}, \citenamefont {Prestage}, \citenamefont {Seubert},\ and\ \citenamefont {Tjoelker}}]{ely2018}%
  \BibitemOpen
  \bibfield  {author} {\bibinfo {author} {\bibfnamefont {T.~A.}\ \bibnamefont {Ely}}, \bibinfo {author} {\bibfnamefont {E.~A.}\ \bibnamefont {Burt}}, \bibinfo {author} {\bibfnamefont {J.~D.}\ \bibnamefont {Prestage}}, \bibinfo {author} {\bibfnamefont {J.~M.}\ \bibnamefont {Seubert}},\ and\ \bibinfo {author} {\bibfnamefont {R.~L.}\ \bibnamefont {Tjoelker}},\ }\bibfield  {title} {\bibinfo {title} {Using the deep space atomic clock for navigation and science},\ }\href {https://doi.org/10.1109/TUFFC.2018.2808269} {\bibfield  {journal} {\bibinfo  {journal} {IEEE Trans. Ultrason. Ferroelectr. Freq. Control}\ }\textbf {\bibinfo {volume} {65}},\ \bibinfo {pages} {950} (\bibinfo {year} {2018})}\BibitemShut {NoStop}%
\bibitem [{\citenamefont {Takamoto}\ \emph {et~al.}(2022)\citenamefont {Takamoto}, \citenamefont {Tanaka},\ and\ \citenamefont {Katori}}]{takamoto2022}%
  \BibitemOpen
  \bibfield  {author} {\bibinfo {author} {\bibfnamefont {M.}~\bibnamefont {Takamoto}}, \bibinfo {author} {\bibfnamefont {Y.}~\bibnamefont {Tanaka}},\ and\ \bibinfo {author} {\bibfnamefont {H.}~\bibnamefont {Katori}},\ }\bibfield  {title} {\bibinfo {title} {{A perspective on the future of transportable optical lattice clocks}},\ }\href {https://doi.org/10.1063/5.0087894} {\bibfield  {journal} {\bibinfo  {journal} {Appl. Phys. Lett.}\ }\textbf {\bibinfo {volume} {120}},\ \bibinfo {pages} {140502} (\bibinfo {year} {2022})}\BibitemShut {NoStop}%
\bibitem [{\citenamefont {Little}\ \emph {et~al.}(2021)\citenamefont {Little}, \citenamefont {Hoth}, \citenamefont {Christensen}, \citenamefont {Walker}, \citenamefont {De~Smet}, \citenamefont {Biedermann}, \citenamefont {Lee},\ and\ \citenamefont {Schwindt}}]{little2021}%
  \BibitemOpen
  \bibfield  {author} {\bibinfo {author} {\bibfnamefont {B.~J.}\ \bibnamefont {Little}}, \bibinfo {author} {\bibfnamefont {G.~W.}\ \bibnamefont {Hoth}}, \bibinfo {author} {\bibfnamefont {J.}~\bibnamefont {Christensen}}, \bibinfo {author} {\bibfnamefont {C.}~\bibnamefont {Walker}}, \bibinfo {author} {\bibfnamefont {D.~J.}\ \bibnamefont {De~Smet}}, \bibinfo {author} {\bibfnamefont {G.~W.}\ \bibnamefont {Biedermann}}, \bibinfo {author} {\bibfnamefont {J.}~\bibnamefont {Lee}},\ and\ \bibinfo {author} {\bibfnamefont {P.~D.~D.}\ \bibnamefont {Schwindt}},\ }\bibfield  {title} {\bibinfo {title} {{A passively pumped vacuum package sustaining cold atoms for more than 200 days}},\ }\href {https://doi.org/10.1116/5.0053885} {\bibfield  {journal} {\bibinfo  {journal} {AVS Quantum Sci.}\ }\textbf {\bibinfo {volume} {3}},\ \bibinfo {pages} {035001} (\bibinfo {year} {2021})}\BibitemShut {NoStop}%
\bibitem [{\citenamefont {McGilligan}\ \emph {et~al.}(2022)\citenamefont {McGilligan}, \citenamefont {Gallacher}, \citenamefont {Griffin}, \citenamefont {Paul}, \citenamefont {Arnold},\ and\ \citenamefont {Riis}}]{mcgilligan2022}%
  \BibitemOpen
  \bibfield  {author} {\bibinfo {author} {\bibfnamefont {J.~P.}\ \bibnamefont {McGilligan}}, \bibinfo {author} {\bibfnamefont {K.}~\bibnamefont {Gallacher}}, \bibinfo {author} {\bibfnamefont {P.~F.}\ \bibnamefont {Griffin}}, \bibinfo {author} {\bibfnamefont {D.~J.}\ \bibnamefont {Paul}}, \bibinfo {author} {\bibfnamefont {A.~S.}\ \bibnamefont {Arnold}},\ and\ \bibinfo {author} {\bibfnamefont {E.}~\bibnamefont {Riis}},\ }\bibfield  {title} {\bibinfo {title} {{Micro-fabricated components for cold atom sensors}},\ }\href {https://doi.org/10.1063/5.0101628} {\bibfield  {journal} {\bibinfo  {journal} {Rev. Sci. Instr.}\ }\textbf {\bibinfo {volume} {93}},\ \bibinfo {pages} {091101} (\bibinfo {year} {2022})}\BibitemShut {NoStop}%
\bibitem [{\citenamefont {Martinez}\ \emph {et~al.}(2023)\citenamefont {Martinez}, \citenamefont {Li}, \citenamefont {Staron}, \citenamefont {Kitching}, \citenamefont {Raman},\ and\ \citenamefont {McGehee}}]{martinez2023}%
  \BibitemOpen
  \bibfield  {author} {\bibinfo {author} {\bibfnamefont {G.~D.}\ \bibnamefont {Martinez}}, \bibinfo {author} {\bibfnamefont {C.}~\bibnamefont {Li}}, \bibinfo {author} {\bibfnamefont {A.}~\bibnamefont {Staron}}, \bibinfo {author} {\bibfnamefont {J.}~\bibnamefont {Kitching}}, \bibinfo {author} {\bibfnamefont {C.}~\bibnamefont {Raman}},\ and\ \bibinfo {author} {\bibfnamefont {W.~R.}\ \bibnamefont {McGehee}},\ }\bibfield  {title} {\bibinfo {title} {A chip-scale atomic beam clock},\ }\href {https://doi.org/10.1038/s41467-023-39166-1} {\bibfield  {journal} {\bibinfo  {journal} {Nat. Comm.}\ }\textbf {\bibinfo {volume} {14}},\ \bibinfo {pages} {3501} (\bibinfo {year} {2023})}\BibitemShut {NoStop}%
\bibitem [{\citenamefont {Bregazzi}\ \emph {et~al.}(2024)\citenamefont {Bregazzi}, \citenamefont {Batori}, \citenamefont {Lewis}, \citenamefont {Affolderbach}, \citenamefont {Mileti}, \citenamefont {Riis},\ and\ \citenamefont {Griffin}}]{bregazzi2023}%
  \BibitemOpen
  \bibfield  {author} {\bibinfo {author} {\bibfnamefont {A.}~\bibnamefont {Bregazzi}}, \bibinfo {author} {\bibfnamefont {E.}~\bibnamefont {Batori}}, \bibinfo {author} {\bibfnamefont {B.}~\bibnamefont {Lewis}}, \bibinfo {author} {\bibfnamefont {C.}~\bibnamefont {Affolderbach}}, \bibinfo {author} {\bibfnamefont {G.}~\bibnamefont {Mileti}}, \bibinfo {author} {\bibfnamefont {E.}~\bibnamefont {Riis}},\ and\ \bibinfo {author} {\bibfnamefont {P.~F.}\ \bibnamefont {Griffin}},\ }\bibfield  {title} {\bibinfo {title} {A cold-atom {R}amsey clock with a low volume physics package},\ }\href {https://doi.org/10.1038/s41598-024-51418-8} {\bibfield  {journal} {\bibinfo  {journal} {Sci. Rep.}\ }\textbf {\bibinfo {volume} {14}},\ \bibinfo {pages} {931} (\bibinfo {year} {2024})}\BibitemShut {NoStop}%
\bibitem [{\citenamefont {Camparo}(2007)}]{camparo2007}%
  \BibitemOpen
  \bibfield  {author} {\bibinfo {author} {\bibfnamefont {J.}~\bibnamefont {Camparo}},\ }\bibfield  {title} {\bibinfo {title} {{The rubidium atomic clock and basic research}},\ }\href {https://doi.org/10.1063/1.2812121} {\bibfield  {journal} {\bibinfo  {journal} {Phys. Today}\ }\textbf {\bibinfo {volume} {60}},\ \bibinfo {pages} {33} (\bibinfo {year} {2007})}\BibitemShut {NoStop}%
\bibitem [{\citenamefont {Martin}\ \emph {et~al.}(2018)\citenamefont {Martin}, \citenamefont {Phelps}, \citenamefont {Lemke}, \citenamefont {Bigelow}, \citenamefont {Stuhl}, \citenamefont {Wojcik}, \citenamefont {Holt}, \citenamefont {Coddington}, \citenamefont {Bishop},\ and\ \citenamefont {Burke}}]{martin2018}%
  \BibitemOpen
  \bibfield  {author} {\bibinfo {author} {\bibfnamefont {K.~W.}\ \bibnamefont {Martin}}, \bibinfo {author} {\bibfnamefont {G.}~\bibnamefont {Phelps}}, \bibinfo {author} {\bibfnamefont {N.~D.}\ \bibnamefont {Lemke}}, \bibinfo {author} {\bibfnamefont {M.~S.}\ \bibnamefont {Bigelow}}, \bibinfo {author} {\bibfnamefont {B.}~\bibnamefont {Stuhl}}, \bibinfo {author} {\bibfnamefont {M.}~\bibnamefont {Wojcik}}, \bibinfo {author} {\bibfnamefont {M.}~\bibnamefont {Holt}}, \bibinfo {author} {\bibfnamefont {I.}~\bibnamefont {Coddington}}, \bibinfo {author} {\bibfnamefont {M.~W.}\ \bibnamefont {Bishop}},\ and\ \bibinfo {author} {\bibfnamefont {J.~H.}\ \bibnamefont {Burke}},\ }\bibfield  {title} {\bibinfo {title} {Compact optical atomic clock based on a two-photon transition in rubidium},\ }\href {https://doi.org/10.1103/PhysRevApplied.9.014019} {\bibfield  {journal} {\bibinfo  {journal} {Phys. Rev. Applied}\ }\textbf {\bibinfo {volume} {9}},\ \bibinfo {pages} {014019} (\bibinfo {year} {2018})}\BibitemShut {NoStop}%
\bibitem [{\citenamefont {Sharma}\ \emph {et~al.}(2022)\citenamefont {Sharma}, \citenamefont {Kolkowitz},\ and\ \citenamefont {Saffman}}]{sharma2022analysis}%
  \BibitemOpen
  \bibfield  {author} {\bibinfo {author} {\bibfnamefont {A.}~\bibnamefont {Sharma}}, \bibinfo {author} {\bibfnamefont {S.}~\bibnamefont {Kolkowitz}},\ and\ \bibinfo {author} {\bibfnamefont {M.}~\bibnamefont {Saffman}},\ }\bibfield  {title} {\bibinfo {title} {Analysis of a cesium lattice optical clock},\ }\href {https://arxiv.org/abs/2203.08708} {\bibfield  {journal} {\bibinfo  {journal} {arXiv:2203.08708}\ } (\bibinfo {year} {2022})}\BibitemShut {NoStop}%
\bibitem [{\citenamefont {Moon}\ \emph {et~al.}(2009)\citenamefont {Moon}, \citenamefont {Lee},\ and\ \citenamefont {Suh}}]{moon2009}%
  \BibitemOpen
  \bibfield  {author} {\bibinfo {author} {\bibfnamefont {H.~S.}\ \bibnamefont {Moon}}, \bibinfo {author} {\bibfnamefont {W.~K.}\ \bibnamefont {Lee}},\ and\ \bibinfo {author} {\bibfnamefont {H.~S.}\ \bibnamefont {Suh}},\ }\bibfield  {title} {\bibinfo {title} {Hyperfine-structure-constant determination and absolute-frequency measurement of the {R}b $4{D}_{3/2}$ state},\ }\href {https://doi.org/10.1103/PhysRevA.79.062503} {\bibfield  {journal} {\bibinfo  {journal} {Phys. Rev. A}\ }\textbf {\bibinfo {volume} {79}},\ \bibinfo {pages} {062503} (\bibinfo {year} {2009})}\BibitemShut {NoStop}%
\bibitem [{\citenamefont {Wang}\ \emph {et~al.}(2014)\citenamefont {Wang}, \citenamefont {Liu}, \citenamefont {Yang}, \citenamefont {Yang},\ and\ \citenamefont {Wang}}]{wang2014}%
  \BibitemOpen
  \bibfield  {author} {\bibinfo {author} {\bibfnamefont {J.}~\bibnamefont {Wang}}, \bibinfo {author} {\bibfnamefont {H.}~\bibnamefont {Liu}}, \bibinfo {author} {\bibfnamefont {G.}~\bibnamefont {Yang}}, \bibinfo {author} {\bibfnamefont {B.}~\bibnamefont {Yang}},\ and\ \bibinfo {author} {\bibfnamefont {J.}~\bibnamefont {Wang}},\ }\bibfield  {title} {\bibinfo {title} {Determination of the hyperfine structure constants of the $^{87}\mathrm{Rb}$ and $^{85}\mathrm{Rb}$ $4{D}_{5/2}$ state and the isotope hyperfine anomaly},\ }\href {https://doi.org/10.1103/PhysRevA.90.052505} {\bibfield  {journal} {\bibinfo  {journal} {Phys. Rev. A}\ }\textbf {\bibinfo {volume} {90}},\ \bibinfo {pages} {052505} (\bibinfo {year} {2014})}\BibitemShut {NoStop}%
\bibitem [{\citenamefont {Lim}\ \emph {et~al.}(2022)\citenamefont {Lim}, \citenamefont {McPoyle},\ and\ \citenamefont {Cervantes}}]{lim2022}%
  \BibitemOpen
  \bibfield  {author} {\bibinfo {author} {\bibfnamefont {M.~J.}\ \bibnamefont {Lim}}, \bibinfo {author} {\bibfnamefont {S.}~\bibnamefont {McPoyle}},\ and\ \bibinfo {author} {\bibfnamefont {M.}~\bibnamefont {Cervantes}},\ }\bibfield  {title} {\bibinfo {title} {Modulation transfer spectroscopy of a four-level ladder system in atomic rubidium},\ }\href {https://doi.org/https://doi.org/10.1016/j.optcom.2022.128651} {\bibfield  {journal} {\bibinfo  {journal} {Opt. Comm.}\ }\textbf {\bibinfo {volume} {522}},\ \bibinfo {pages} {128651} (\bibinfo {year} {2022})}\BibitemShut {NoStop}%
\bibitem [{\citenamefont {Duspayev}\ and\ \citenamefont {Raithel}(2023)}]{Duspayev2023}%
  \BibitemOpen
  \bibfield  {author} {\bibinfo {author} {\bibfnamefont {A.}~\bibnamefont {Duspayev}}\ and\ \bibinfo {author} {\bibfnamefont {G.}~\bibnamefont {Raithel}},\ }\bibfield  {title} {\bibinfo {title} {Spectroscopy of the $^{85}${R}b 4${D}_{3/2}$ state for hyperfine-structure determination},\ }\href {https://doi.org/10.1088/1367-2630/acf405} {\bibfield  {journal} {\bibinfo  {journal} {New J. Phys.}\ }\textbf {\bibinfo {volume} {25}},\ \bibinfo {pages} {093015} (\bibinfo {year} {2023})}\BibitemShut {NoStop}%
\bibitem [{\citenamefont {Chaneli\`ere}\ \emph {et~al.}(2006)\citenamefont {Chaneli\`ere}, \citenamefont {Matsukevich}, \citenamefont {Jenkins}, \citenamefont {Kennedy}, \citenamefont {Chapman},\ and\ \citenamefont {Kuzmich}}]{chanliere2006}%
  \BibitemOpen
  \bibfield  {author} {\bibinfo {author} {\bibfnamefont {T.}~\bibnamefont {Chaneli\`ere}}, \bibinfo {author} {\bibfnamefont {D.~N.}\ \bibnamefont {Matsukevich}}, \bibinfo {author} {\bibfnamefont {S.~D.}\ \bibnamefont {Jenkins}}, \bibinfo {author} {\bibfnamefont {T.~A.~B.}\ \bibnamefont {Kennedy}}, \bibinfo {author} {\bibfnamefont {M.~S.}\ \bibnamefont {Chapman}},\ and\ \bibinfo {author} {\bibfnamefont {A.}~\bibnamefont {Kuzmich}},\ }\bibfield  {title} {\bibinfo {title} {Quantum telecommunication based on atomic cascade transitions},\ }\href {https://doi.org/10.1103/PhysRevLett.96.093604} {\bibfield  {journal} {\bibinfo  {journal} {Phys. Rev. Lett.}\ }\textbf {\bibinfo {volume} {96}},\ \bibinfo {pages} {093604} (\bibinfo {year} {2006})}\BibitemShut {NoStop}%
\bibitem [{\citenamefont {Huie}\ \emph {et~al.}(2021)\citenamefont {Huie}, \citenamefont {Menon}, \citenamefont {Bernien},\ and\ \citenamefont {Covey}}]{huie2021}%
  \BibitemOpen
  \bibfield  {author} {\bibinfo {author} {\bibfnamefont {W.}~\bibnamefont {Huie}}, \bibinfo {author} {\bibfnamefont {S.~G.}\ \bibnamefont {Menon}}, \bibinfo {author} {\bibfnamefont {H.}~\bibnamefont {Bernien}},\ and\ \bibinfo {author} {\bibfnamefont {J.~P.}\ \bibnamefont {Covey}},\ }\bibfield  {title} {\bibinfo {title} {Multiplexed telecommunication-band quantum networking with atom arrays in optical cavities},\ }\href {https://doi.org/10.1103/PhysRevResearch.3.043154} {\bibfield  {journal} {\bibinfo  {journal} {Phys. Rev. Res.}\ }\textbf {\bibinfo {volume} {3}},\ \bibinfo {pages} {043154} (\bibinfo {year} {2021})}\BibitemShut {NoStop}%
\bibitem [{\citenamefont {K\'om\'ar}\ \emph {et~al.}(2016)\citenamefont {K\'om\'ar}, \citenamefont {Topcu}, \citenamefont {Kessler}, \citenamefont {Derevianko}, \citenamefont {Vuleti\ifmmode~\acute{c}\else \'{c}\fi{}}, \citenamefont {Ye},\ and\ \citenamefont {Lukin}}]{komar2016}%
  \BibitemOpen
  \bibfield  {author} {\bibinfo {author} {\bibfnamefont {P.}~\bibnamefont {K\'om\'ar}}, \bibinfo {author} {\bibfnamefont {T.}~\bibnamefont {Topcu}}, \bibinfo {author} {\bibfnamefont {E.~M.}\ \bibnamefont {Kessler}}, \bibinfo {author} {\bibfnamefont {A.}~\bibnamefont {Derevianko}}, \bibinfo {author} {\bibfnamefont {V.}~\bibnamefont {Vuleti\ifmmode~\acute{c}\else \'{c}\fi{}}}, \bibinfo {author} {\bibfnamefont {J.}~\bibnamefont {Ye}},\ and\ \bibinfo {author} {\bibfnamefont {M.~D.}\ \bibnamefont {Lukin}},\ }\bibfield  {title} {\bibinfo {title} {Quantum network of atom clocks: A possible implementation with neutral atoms},\ }\href {https://doi.org/10.1103/PhysRevLett.117.060506} {\bibfield  {journal} {\bibinfo  {journal} {Phys. Rev. Lett.}\ }\textbf {\bibinfo {volume} {117}},\ \bibinfo {pages} {060506} (\bibinfo {year} {2016})}\BibitemShut {NoStop}%
\bibitem [{\citenamefont {{Boulder Atomic Clock Optical Network (BACON) Collaboration}*}(2021)}]{BACON2021}%
  \BibitemOpen
  \bibfield  {author} {\bibinfo {author} {\bibnamefont {{Boulder Atomic Clock Optical Network (BACON) Collaboration}*}},\ }\bibfield  {title} {\bibinfo {title} {Frequency ratio measurements at 18-digit accuracy using an optical clock network},\ }\href {https://doi.org/10.1038/s41586-021-03253-4} {\bibfield  {journal} {\bibinfo  {journal} {Nature}\ }\textbf {\bibinfo {volume} {591}},\ \bibinfo {pages} {564} (\bibinfo {year} {2021})}\BibitemShut {NoStop}%
\bibitem [{\citenamefont {Nichol}\ \emph {et~al.}(2022)\citenamefont {Nichol}, \citenamefont {Srinivas}, \citenamefont {Nadlinger}, \citenamefont {Drmota}, \citenamefont {Main}, \citenamefont {Araneda}, \citenamefont {Ballance},\ and\ \citenamefont {Lucas}}]{nichol2022}%
  \BibitemOpen
  \bibfield  {author} {\bibinfo {author} {\bibfnamefont {B.~C.}\ \bibnamefont {Nichol}}, \bibinfo {author} {\bibfnamefont {R.}~\bibnamefont {Srinivas}}, \bibinfo {author} {\bibfnamefont {D.~P.}\ \bibnamefont {Nadlinger}}, \bibinfo {author} {\bibfnamefont {P.}~\bibnamefont {Drmota}}, \bibinfo {author} {\bibfnamefont {D.}~\bibnamefont {Main}}, \bibinfo {author} {\bibfnamefont {G.}~\bibnamefont {Araneda}}, \bibinfo {author} {\bibfnamefont {C.~J.}\ \bibnamefont {Ballance}},\ and\ \bibinfo {author} {\bibfnamefont {D.~M.}\ \bibnamefont {Lucas}},\ }\bibfield  {title} {\bibinfo {title} {An elementary quantum network of entangled optical atomic clocks},\ }\href {https://doi.org/10.1038/s41586-022-05088-z} {\bibfield  {journal} {\bibinfo  {journal} {Nature}\ }\textbf {\bibinfo {volume} {609}},\ \bibinfo {pages} {689} (\bibinfo {year} {2022})}\BibitemShut {NoStop}%
\bibitem [{\citenamefont {Eustice}\ \emph {et~al.}(2023)\citenamefont {Eustice}, \citenamefont {Filin}, \citenamefont {Schrott}, \citenamefont {Porsev}, \citenamefont {Cheung}, \citenamefont {Novoa}, \citenamefont {Stamper-Kurn},\ and\ \citenamefont {Safronova}}]{eustice2023}%
  \BibitemOpen
  \bibfield  {author} {\bibinfo {author} {\bibfnamefont {S.}~\bibnamefont {Eustice}}, \bibinfo {author} {\bibfnamefont {D.}~\bibnamefont {Filin}}, \bibinfo {author} {\bibfnamefont {J.}~\bibnamefont {Schrott}}, \bibinfo {author} {\bibfnamefont {S.}~\bibnamefont {Porsev}}, \bibinfo {author} {\bibfnamefont {C.}~\bibnamefont {Cheung}}, \bibinfo {author} {\bibfnamefont {D.}~\bibnamefont {Novoa}}, \bibinfo {author} {\bibfnamefont {D.~M.}\ \bibnamefont {Stamper-Kurn}},\ and\ \bibinfo {author} {\bibfnamefont {M.~S.}\ \bibnamefont {Safronova}},\ }\bibfield  {title} {\bibinfo {title} {Optical telecommunications-band clock based on neutral titanium atoms},\ }\href {https://doi.org/10.1103/PhysRevA.107.L051102} {\bibfield  {journal} {\bibinfo  {journal} {Phys. Rev. A}\ }\textbf {\bibinfo {volume} {107}},\ \bibinfo {pages} {L051102} (\bibinfo {year} {2023})}\BibitemShut {NoStop}%
\bibitem [{\citenamefont {Duspayev}\ \emph {et~al.}(2024)\citenamefont {Duspayev}, \citenamefont {Cardman}, \citenamefont {Anderson},\ and\ \citenamefont {Raithel}}]{duspayev2023highell}%
  \BibitemOpen
  \bibfield  {author} {\bibinfo {author} {\bibfnamefont {A.}~\bibnamefont {Duspayev}}, \bibinfo {author} {\bibfnamefont {R.}~\bibnamefont {Cardman}}, \bibinfo {author} {\bibfnamefont {D.~A.}\ \bibnamefont {Anderson}},\ and\ \bibinfo {author} {\bibfnamefont {G.}~\bibnamefont {Raithel}},\ }\bibfield  {title} {\bibinfo {title} {High-angular-momentum {R}ydberg states in a room-temperature vapor cell for dc electric-field sensing},\ }\href {https://doi.org/10.1103/PhysRevResearch.6.023138} {\bibfield  {journal} {\bibinfo  {journal} {Phys. Rev. Res.}\ }\textbf {\bibinfo {volume} {6}},\ \bibinfo {pages} {023138} (\bibinfo {year} {2024})}\BibitemShut {NoStop}%
\bibitem [{\citenamefont {Shaffer}\ \emph {et~al.}(2018)\citenamefont {Shaffer}, \citenamefont {Rittenhouse},\ and\ \citenamefont {Sadeghpour}}]{shafferreview}%
  \BibitemOpen
  \bibfield  {author} {\bibinfo {author} {\bibfnamefont {J.~P.}\ \bibnamefont {Shaffer}}, \bibinfo {author} {\bibfnamefont {S.~T.}\ \bibnamefont {Rittenhouse}},\ and\ \bibinfo {author} {\bibfnamefont {H.~R.}\ \bibnamefont {Sadeghpour}},\ }\bibfield  {title} {\bibinfo {title} {Ultracold {R}ydberg molecules},\ }\href {https://doi.org/10.1038/s41467-018-04135-6} {\bibfield  {journal} {\bibinfo  {journal} {Nat. Comm.}\ }\textbf {\bibinfo {volume} {9}},\ \bibinfo {pages} {1965} (\bibinfo {year} {2018})}\BibitemShut {NoStop}%
\bibitem [{\citenamefont {Duspayev}\ \emph {et~al.}(2021)\citenamefont {Duspayev}, \citenamefont {Han}, \citenamefont {Viray}, \citenamefont {Ma}, \citenamefont {Zhao},\ and\ \citenamefont {Raithel}}]{duspayev2021}%
  \BibitemOpen
  \bibfield  {author} {\bibinfo {author} {\bibfnamefont {A.}~\bibnamefont {Duspayev}}, \bibinfo {author} {\bibfnamefont {X.}~\bibnamefont {Han}}, \bibinfo {author} {\bibfnamefont {M.~A.}\ \bibnamefont {Viray}}, \bibinfo {author} {\bibfnamefont {L.}~\bibnamefont {Ma}}, \bibinfo {author} {\bibfnamefont {J.}~\bibnamefont {Zhao}},\ and\ \bibinfo {author} {\bibfnamefont {G.}~\bibnamefont {Raithel}},\ }\bibfield  {title} {\bibinfo {title} {Long-range {R}ydberg-atom--ion molecules of {R}b and {C}s},\ }\href {https://doi.org/10.1103/PhysRevResearch.3.023114} {\bibfield  {journal} {\bibinfo  {journal} {Phys. Rev. Research}\ }\textbf {\bibinfo {volume} {3}},\ \bibinfo {pages} {023114} (\bibinfo {year} {2021})}\BibitemShut {NoStop}%
\bibitem [{\citenamefont {Cardman}\ and\ \citenamefont {Raithel}(2020)}]{cardman2020}%
  \BibitemOpen
  \bibfield  {author} {\bibinfo {author} {\bibfnamefont {R.}~\bibnamefont {Cardman}}\ and\ \bibinfo {author} {\bibfnamefont {G.}~\bibnamefont {Raithel}},\ }\bibfield  {title} {\bibinfo {title} {Circularizing {R}ydberg atoms with time-dependent optical traps},\ }\href {https://doi.org/10.1103/PhysRevA.101.013434} {\bibfield  {journal} {\bibinfo  {journal} {Phys. Rev. A}\ }\textbf {\bibinfo {volume} {101}},\ \bibinfo {pages} {013434} (\bibinfo {year} {2020})}\BibitemShut {NoStop}%
\bibitem [{\citenamefont {Martin}\ \emph {et~al.}(2019)\citenamefont {Martin}, \citenamefont {Stuhl}, \citenamefont {Eugenio}, \citenamefont {Safronova}, \citenamefont {Phelps}, \citenamefont {Burke},\ and\ \citenamefont {Lemke}}]{martinpra2019}%
  \BibitemOpen
  \bibfield  {author} {\bibinfo {author} {\bibfnamefont {K.~W.}\ \bibnamefont {Martin}}, \bibinfo {author} {\bibfnamefont {B.}~\bibnamefont {Stuhl}}, \bibinfo {author} {\bibfnamefont {J.}~\bibnamefont {Eugenio}}, \bibinfo {author} {\bibfnamefont {M.~S.}\ \bibnamefont {Safronova}}, \bibinfo {author} {\bibfnamefont {G.}~\bibnamefont {Phelps}}, \bibinfo {author} {\bibfnamefont {J.~H.}\ \bibnamefont {Burke}},\ and\ \bibinfo {author} {\bibfnamefont {N.~D.}\ \bibnamefont {Lemke}},\ }\bibfield  {title} {\bibinfo {title} {Frequency shifts due to stark effects on a rubidium two-photon transition},\ }\href {https://doi.org/10.1103/PhysRevA.100.023417} {\bibfield  {journal} {\bibinfo  {journal} {Phys. Rev. A}\ }\textbf {\bibinfo {volume} {100}},\ \bibinfo {pages} {023417} (\bibinfo {year} {2019})}\BibitemShut {NoStop}%
\bibitem [{\citenamefont {Gerginov}\ and\ \citenamefont {Beloy}(2018)}]{gerginov2018}%
  \BibitemOpen
  \bibfield  {author} {\bibinfo {author} {\bibfnamefont {V.}~\bibnamefont {Gerginov}}\ and\ \bibinfo {author} {\bibfnamefont {K.}~\bibnamefont {Beloy}},\ }\bibfield  {title} {\bibinfo {title} {Two-photon optical frequency reference with active ac {S}tark shift cancellation},\ }\href {https://doi.org/10.1103/PhysRevApplied.10.014031} {\bibfield  {journal} {\bibinfo  {journal} {Phys. Rev. Appl.}\ }\textbf {\bibinfo {volume} {10}},\ \bibinfo {pages} {014031} (\bibinfo {year} {2018})}\BibitemShut {NoStop}%
\bibitem [{\citenamefont {Perrella}\ \emph {et~al.}(2019)\citenamefont {Perrella}, \citenamefont {Light}, \citenamefont {Anstie}, \citenamefont {Baynes}, \citenamefont {White},\ and\ \citenamefont {Luiten}}]{perrella2019}%
  \BibitemOpen
  \bibfield  {author} {\bibinfo {author} {\bibfnamefont {C.}~\bibnamefont {Perrella}}, \bibinfo {author} {\bibfnamefont {P.}~\bibnamefont {Light}}, \bibinfo {author} {\bibfnamefont {J.}~\bibnamefont {Anstie}}, \bibinfo {author} {\bibfnamefont {F.}~\bibnamefont {Baynes}}, \bibinfo {author} {\bibfnamefont {R.}~\bibnamefont {White}},\ and\ \bibinfo {author} {\bibfnamefont {A.}~\bibnamefont {Luiten}},\ }\bibfield  {title} {\bibinfo {title} {Dichroic two-photon rubidium frequency standard},\ }\href {https://doi.org/10.1103/PhysRevApplied.12.054063} {\bibfield  {journal} {\bibinfo  {journal} {Phys. Rev. Appl.}\ }\textbf {\bibinfo {volume} {12}},\ \bibinfo {pages} {054063} (\bibinfo {year} {2019})}\BibitemShut {NoStop}%
\bibitem [{\citenamefont {Cardman}\ \emph {et~al.}(2021)\citenamefont {Cardman}, \citenamefont {Han}, \citenamefont {MacLennan}, \citenamefont {Duspayev},\ and\ \citenamefont {Raithel}}]{cardman2021}%
  \BibitemOpen
  \bibfield  {author} {\bibinfo {author} {\bibfnamefont {R.}~\bibnamefont {Cardman}}, \bibinfo {author} {\bibfnamefont {X.}~\bibnamefont {Han}}, \bibinfo {author} {\bibfnamefont {J.~L.}\ \bibnamefont {MacLennan}}, \bibinfo {author} {\bibfnamefont {A.}~\bibnamefont {Duspayev}},\ and\ \bibinfo {author} {\bibfnamefont {G.}~\bibnamefont {Raithel}},\ }\bibfield  {title} {\bibinfo {title} {ac polarizability and photoionization-cross-section measurements in an optical lattice},\ }\href {https://doi.org/10.1103/PhysRevA.104.063304} {\bibfield  {journal} {\bibinfo  {journal} {Phys. Rev. A}\ }\textbf {\bibinfo {volume} {104}},\ \bibinfo {pages} {063304} (\bibinfo {year} {2021})}\BibitemShut {NoStop}%
\bibitem [{\citenamefont {Duspayev}\ \emph {et~al.}(2022)\citenamefont {Duspayev}, \citenamefont {Cardman},\ and\ \citenamefont {Raithel}}]{atoms10040117}%
  \BibitemOpen
  \bibfield  {author} {\bibinfo {author} {\bibfnamefont {A.}~\bibnamefont {Duspayev}}, \bibinfo {author} {\bibfnamefont {R.}~\bibnamefont {Cardman}},\ and\ \bibinfo {author} {\bibfnamefont {G.}~\bibnamefont {Raithel}},\ }\bibfield  {title} {\bibinfo {title} {Dynamic polarizability of the $^{85}${R}b 5{D}$_{3/2}$-state in 1064 nm light},\ }\href {https://doi.org/10.3390/atoms10040117} {\bibfield  {journal} {\bibinfo  {journal} {Atoms}\ }\textbf {\bibinfo {volume} {10}},\ \bibinfo {pages} {117} (\bibinfo {year} {2022})}\BibitemShut {NoStop}%
\bibitem [{\citenamefont {Safronova}\ \emph {et~al.}(2004)\citenamefont {Safronova}, \citenamefont {Williams},\ and\ \citenamefont {Clark}}]{safronova2004}%
  \BibitemOpen
  \bibfield  {author} {\bibinfo {author} {\bibfnamefont {M.~S.}\ \bibnamefont {Safronova}}, \bibinfo {author} {\bibfnamefont {C.~J.}\ \bibnamefont {Williams}},\ and\ \bibinfo {author} {\bibfnamefont {C.~W.}\ \bibnamefont {Clark}},\ }\bibfield  {title} {\bibinfo {title} {Relativistic many-body calculations of electric-dipole matrix elements, lifetimes, and polarizabilities in rubidium},\ }\href {https://doi.org/10.1103/PhysRevA.69.022509} {\bibfield  {journal} {\bibinfo  {journal} {Phys. Rev. A}\ }\textbf {\bibinfo {volume} {69}},\ \bibinfo {pages} {022509} (\bibinfo {year} {2004})}\BibitemShut {NoStop}%
\bibitem [{\citenamefont {Safronova}\ and\ \citenamefont {Safronova}(2011)}]{safronova2011}%
  \BibitemOpen
  \bibfield  {author} {\bibinfo {author} {\bibfnamefont {M.~S.}\ \bibnamefont {Safronova}}\ and\ \bibinfo {author} {\bibfnamefont {U.~I.}\ \bibnamefont {Safronova}},\ }\bibfield  {title} {\bibinfo {title} {Critically evaluated theoretical energies, lifetimes, hyperfine constants, and multipole polarizabilities in $^{87}\mathrm{{R}b}$},\ }\href {https://doi.org/10.1103/PhysRevA.83.052508} {\bibfield  {journal} {\bibinfo  {journal} {Phys. Rev. A}\ }\textbf {\bibinfo {volume} {83}},\ \bibinfo {pages} {052508} (\bibinfo {year} {2011})}\BibitemShut {NoStop}%
\bibitem [{\citenamefont {Reinhard}\ \emph {et~al.}(2007)\citenamefont {Reinhard}, \citenamefont {Liebisch}, \citenamefont {Knuffman},\ and\ \citenamefont {Raithel}}]{reinhardtA2007}%
  \BibitemOpen
  \bibfield  {author} {\bibinfo {author} {\bibfnamefont {A.}~\bibnamefont {Reinhard}}, \bibinfo {author} {\bibfnamefont {T.~C.}\ \bibnamefont {Liebisch}}, \bibinfo {author} {\bibfnamefont {B.}~\bibnamefont {Knuffman}},\ and\ \bibinfo {author} {\bibfnamefont {G.}~\bibnamefont {Raithel}},\ }\bibfield  {title} {\bibinfo {title} {Level shifts of rubidium {R}ydberg states due to binary interactions},\ }\href {https://doi.org/10.1103/PhysRevA.75.032712} {\bibfield  {journal} {\bibinfo  {journal} {Phys. Rev. A}\ }\textbf {\bibinfo {volume} {75}},\ \bibinfo {pages} {032712} (\bibinfo {year} {2007})}\BibitemShut {NoStop}%
\bibitem [{\citenamefont {Marinescu}\ \emph {et~al.}(1994)\citenamefont {Marinescu}, \citenamefont {Sadeghpour},\ and\ \citenamefont {Dalgarno}}]{Marinescu1994}%
  \BibitemOpen
  \bibfield  {author} {\bibinfo {author} {\bibfnamefont {M.}~\bibnamefont {Marinescu}}, \bibinfo {author} {\bibfnamefont {H.~R.}\ \bibnamefont {Sadeghpour}},\ and\ \bibinfo {author} {\bibfnamefont {A.}~\bibnamefont {Dalgarno}},\ }\bibfield  {title} {\bibinfo {title} {Dispersion coefficients for alkali-metal dimers},\ }\href {https://doi.org/10.1103/PhysRevA.49.982} {\bibfield  {journal} {\bibinfo  {journal} {Phys. Rev. A}\ }\textbf {\bibinfo {volume} {49}},\ \bibinfo {pages} {982} (\bibinfo {year} {1994})}\BibitemShut {NoStop}%
\bibitem [{\citenamefont {Roy}\ \emph {et~al.}(2017)\citenamefont {Roy}, \citenamefont {Condylis}, \citenamefont {Johnathan},\ and\ \citenamefont {Hessmo}}]{Roy2017}%
  \BibitemOpen
  \bibfield  {author} {\bibinfo {author} {\bibfnamefont {R.}~\bibnamefont {Roy}}, \bibinfo {author} {\bibfnamefont {P.~C.}\ \bibnamefont {Condylis}}, \bibinfo {author} {\bibfnamefont {Y.~J.}\ \bibnamefont {Johnathan}},\ and\ \bibinfo {author} {\bibfnamefont {B.}~\bibnamefont {Hessmo}},\ }\bibfield  {title} {\bibinfo {title} {Atomic frequency reference at 1033 nm for ytterbium ({Y}b)-doped fiber lasers and applications exploiting a rubidium ({R}b) 5$s_{1/2}$ to 4$d_{5/2}$ one-colour two-photon transition},\ }\href {https://doi.org/10.1364/OE.25.007960} {\bibfield  {journal} {\bibinfo  {journal} {Opt. Express}\ }\textbf {\bibinfo {volume} {25}},\ \bibinfo {pages} {7960} (\bibinfo {year} {2017})}\BibitemShut {NoStop}%
\bibitem [{\citenamefont {Sheng}\ \emph {et~al.}(2008)\citenamefont {Sheng}, \citenamefont {P\'erez~Galv\'an},\ and\ \citenamefont {Orozco}}]{shengpra2008}%
  \BibitemOpen
  \bibfield  {author} {\bibinfo {author} {\bibfnamefont {D.}~\bibnamefont {Sheng}}, \bibinfo {author} {\bibfnamefont {A.}~\bibnamefont {P\'erez~Galv\'an}},\ and\ \bibinfo {author} {\bibfnamefont {L.~A.}\ \bibnamefont {Orozco}},\ }\bibfield  {title} {\bibinfo {title} {Lifetime measurements of the $5d$ states of rubidium},\ }\href {https://doi.org/10.1103/PhysRevA.78.062506} {\bibfield  {journal} {\bibinfo  {journal} {Phys. Rev. A}\ }\textbf {\bibinfo {volume} {78}},\ \bibinfo {pages} {062506} (\bibinfo {year} {2008})}\BibitemShut {NoStop}%
\bibitem [{\citenamefont {Heavens}(1961)}]{Heavens_61}%
  \BibitemOpen
  \bibfield  {author} {\bibinfo {author} {\bibfnamefont {O.~S.}\ \bibnamefont {Heavens}},\ }\bibfield  {title} {\bibinfo {title} {Radiative transition probabilities of the lower excited states of the alkali metals},\ }\href {https://doi.org/10.1364/JOSA.51.001058} {\bibfield  {journal} {\bibinfo  {journal} {J. Opt. Soc. Am.}\ }\textbf {\bibinfo {volume} {51}},\ \bibinfo {pages} {1058} (\bibinfo {year} {1961})}\BibitemShut {NoStop}%
\bibitem [{\citenamefont {Hamilton}\ \emph {et~al.}(2023)\citenamefont {Hamilton}, \citenamefont {Roberts}, \citenamefont {Scholten}, \citenamefont {Locke}, \citenamefont {Luiten}, \citenamefont {Ginges},\ and\ \citenamefont {Perrella}}]{hamilton2023}%
  \BibitemOpen
  \bibfield  {author} {\bibinfo {author} {\bibfnamefont {R.}~\bibnamefont {Hamilton}}, \bibinfo {author} {\bibfnamefont {B.~M.}\ \bibnamefont {Roberts}}, \bibinfo {author} {\bibfnamefont {S.~K.}\ \bibnamefont {Scholten}}, \bibinfo {author} {\bibfnamefont {C.}~\bibnamefont {Locke}}, \bibinfo {author} {\bibfnamefont {A.~N.}\ \bibnamefont {Luiten}}, \bibinfo {author} {\bibfnamefont {J.~S.}\ \bibnamefont {Ginges}},\ and\ \bibinfo {author} {\bibfnamefont {C.}~\bibnamefont {Perrella}},\ }\bibfield  {title} {\bibinfo {title} {Experimental and theoretical study of dynamic polarizabilities in the $5{S}_{1/2}$--$5{D}_{5/2}$ clock transition in {R}ubidium-87 and determination of electric dipole matrix elements},\ }\href {https://doi.org/10.1103/PhysRevApplied.19.054059} {\bibfield  {journal} {\bibinfo  {journal} {Phys. Rev. Appl.}\ }\textbf {\bibinfo {volume} {19}},\ \bibinfo {pages} {054059} (\bibinfo {year} {2023})}\BibitemShut {NoStop}%
\bibitem [{\citenamefont {Tjoelker}\ \emph {et~al.}(2016)\citenamefont {Tjoelker} \emph {et~al.}}]{Tjoelker2016}%
  \BibitemOpen
  \bibfield  {author} {\bibinfo {author} {\bibfnamefont {R.~L.}\ \bibnamefont {Tjoelker}} \emph {et~al.},\ }\bibfield  {title} {\bibinfo {title} {Mercury ion clock for a {NASA} technology demonstration mission},\ }\href {https://doi.org/10.1109/TUFFC.2016.2543738} {\bibfield  {journal} {\bibinfo  {journal} {IEEE Trans. Ultrason. Ferroelectr. Freq. Control}\ }\textbf {\bibinfo {volume} {63}},\ \bibinfo {pages} {1034} (\bibinfo {year} {2016})}\BibitemShut {NoStop}%
\bibitem [{\citenamefont {Enzer}\ \emph {et~al.}(2021)\citenamefont {Enzer}, \citenamefont {Murphy},\ and\ \citenamefont {Burt}}]{Enzer2021}%
  \BibitemOpen
  \bibfield  {author} {\bibinfo {author} {\bibfnamefont {D.~G.}\ \bibnamefont {Enzer}}, \bibinfo {author} {\bibfnamefont {D.~W.}\ \bibnamefont {Murphy}},\ and\ \bibinfo {author} {\bibfnamefont {E.~A.}\ \bibnamefont {Burt}},\ }\bibfield  {title} {\bibinfo {title} {Allan deviation of atomic clock frequency corrections: A new diagnostic tool for characterizing clock disturbances},\ }\href {https://doi.org/10.1109/TUFFC.2021.3061005} {\bibfield  {journal} {\bibinfo  {journal} {IEEE Trans. Ultrason. Ferroelectr. Freq. Control}\ }\textbf {\bibinfo {volume} {68}},\ \bibinfo {pages} {2590} (\bibinfo {year} {2021})}\BibitemShut {NoStop}%
\bibitem [{\citenamefont {Ramsey}(1950)}]{Ramsey1950}%
  \BibitemOpen
  \bibfield  {author} {\bibinfo {author} {\bibfnamefont {N.~F.}\ \bibnamefont {Ramsey}},\ }\bibfield  {title} {\bibinfo {title} {A molecular beam resonance method with separated oscillating fields},\ }\href {https://doi.org/10.1103/PhysRev.78.695} {\bibfield  {journal} {\bibinfo  {journal} {Phys. Rev.}\ }\textbf {\bibinfo {volume} {78}},\ \bibinfo {pages} {695} (\bibinfo {year} {1950})}\BibitemShut {NoStop}%
\bibitem [{\citenamefont {Itano}\ \emph {et~al.}(1993)\citenamefont {Itano}, \citenamefont {Bergquist}, \citenamefont {Bollinger}, \citenamefont {Gilligan}, \citenamefont {Heinzen}, \citenamefont {Moore}, \citenamefont {Raizen},\ and\ \citenamefont {Wineland}}]{Itano1993}%
  \BibitemOpen
  \bibfield  {author} {\bibinfo {author} {\bibfnamefont {W.~M.}\ \bibnamefont {Itano}}, \bibinfo {author} {\bibfnamefont {J.~C.}\ \bibnamefont {Bergquist}}, \bibinfo {author} {\bibfnamefont {J.~J.}\ \bibnamefont {Bollinger}}, \bibinfo {author} {\bibfnamefont {J.~M.}\ \bibnamefont {Gilligan}}, \bibinfo {author} {\bibfnamefont {D.~J.}\ \bibnamefont {Heinzen}}, \bibinfo {author} {\bibfnamefont {F.~L.}\ \bibnamefont {Moore}}, \bibinfo {author} {\bibfnamefont {M.~G.}\ \bibnamefont {Raizen}},\ and\ \bibinfo {author} {\bibfnamefont {D.~J.}\ \bibnamefont {Wineland}},\ }\bibfield  {title} {\bibinfo {title} {Quantum projection noise: Population fluctuations in two-level systems},\ }\href {https://doi.org/10.1103/PhysRevA.47.3554} {\bibfield  {journal} {\bibinfo  {journal} {Phys. Rev. A}\ }\textbf {\bibinfo {volume} {47}},\ \bibinfo {pages} {3554} (\bibinfo {year} {1993})}\BibitemShut {NoStop}%
\bibitem [{\citenamefont {Arora}\ and\ \citenamefont {Sahoo}(2012)}]{Arora2012}%
  \BibitemOpen
  \bibfield  {author} {\bibinfo {author} {\bibfnamefont {B.}~\bibnamefont {Arora}}\ and\ \bibinfo {author} {\bibfnamefont {B.~K.}\ \bibnamefont {Sahoo}},\ }\bibfield  {title} {\bibinfo {title} {State-insensitive trapping of {R}b atoms: Linearly versus circularly polarized light},\ }\href {https://doi.org/10.1103/PhysRevA.86.033416} {\bibfield  {journal} {\bibinfo  {journal} {Phys. Rev. A}\ }\textbf {\bibinfo {volume} {86}},\ \bibinfo {pages} {033416} (\bibinfo {year} {2012})}\BibitemShut {NoStop}%
\bibitem [{\citenamefont {Metcalf}\ and\ \citenamefont {van~der Straten}(1999)}]{Metcalf}%
  \BibitemOpen
  \bibfield  {author} {\bibinfo {author} {\bibfnamefont {H.}~\bibnamefont {Metcalf}}\ and\ \bibinfo {author} {\bibfnamefont {P.}~\bibnamefont {van~der Straten}},\ }\href@noop {} {\emph {\bibinfo {title} {Laser Cooling and Trapping}}},\ Vol.~\bibinfo {volume} {3}\ (\bibinfo  {publisher} {Springer-Verlag, New York},\ \bibinfo {year} {1999})\BibitemShut {NoStop}%
\bibitem [{\citenamefont {Nez}\ \emph {et~al.}(1993)\citenamefont {Nez}, \citenamefont {Biraben}, \citenamefont {Felder},\ and\ \citenamefont {Millerioux}}]{Nez1993}%
  \BibitemOpen
  \bibfield  {author} {\bibinfo {author} {\bibfnamefont {F.}~\bibnamefont {Nez}}, \bibinfo {author} {\bibfnamefont {F.}~\bibnamefont {Biraben}}, \bibinfo {author} {\bibfnamefont {R.}~\bibnamefont {Felder}},\ and\ \bibinfo {author} {\bibfnamefont {Y.}~\bibnamefont {Millerioux}},\ }\bibfield  {title} {\bibinfo {title} {Optical frequency determination of the hyperfine components of the 5s12-5d32 two-photon transitions in rubidium},\ }\href {https://doi.org/https://doi.org/10.1016/0030-4018(93)90417-4} {\bibfield  {journal} {\bibinfo  {journal} {Opt. Comm.}\ }\textbf {\bibinfo {volume} {102}},\ \bibinfo {pages} {432} (\bibinfo {year} {1993})}\BibitemShut {NoStop}%
\bibitem [{\citenamefont {Barakhshan}\ \emph {et~al.}()\citenamefont {Barakhshan}, \citenamefont {Marrs}, \citenamefont {Bhosale}, \citenamefont {Arora}, \citenamefont {Eigenmann},\ and\ \citenamefont {Safronova}}]{UDportal}%
  \BibitemOpen
  \bibfield  {author} {\bibinfo {author} {\bibfnamefont {P.}~\bibnamefont {Barakhshan}}, \bibinfo {author} {\bibfnamefont {A.}~\bibnamefont {Marrs}}, \bibinfo {author} {\bibfnamefont {A.}~\bibnamefont {Bhosale}}, \bibinfo {author} {\bibfnamefont {B.}~\bibnamefont {Arora}}, \bibinfo {author} {\bibfnamefont {R.}~\bibnamefont {Eigenmann}},\ and\ \bibinfo {author} {\bibfnamefont {M.~S.}\ \bibnamefont {Safronova}},\ }\href@noop {} {}\bibinfo {howpublished} {\textit{Portal for High-Precision Atomic Data and Computation} (version 2.0). University of Delaware, Newark, DE, USA. \url{https://www.udel.edu/atom} [February 2022]}\BibitemShut {NoStop}%
\bibitem [{\citenamefont {Safronova}\ \emph {et~al.}(2012)\citenamefont {Safronova}, \citenamefont {Kozlov},\ and\ \citenamefont {Clark}}]{safronova_ieee_2012}%
  \BibitemOpen
  \bibfield  {author} {\bibinfo {author} {\bibfnamefont {M.~S.}\ \bibnamefont {Safronova}}, \bibinfo {author} {\bibfnamefont {M.~G.}\ \bibnamefont {Kozlov}},\ and\ \bibinfo {author} {\bibfnamefont {C.~W.}\ \bibnamefont {Clark}},\ }\bibfield  {title} {\bibinfo {title} {Blackbody radiation shifts in optical atomic clocks},\ }\href {https://doi.org/10.1109/TUFFC.2012.2213} {\bibfield  {journal} {\bibinfo  {journal} {IEEE Trans. Ultrason. Ferroelectr. Freq.}\ }\textbf {\bibinfo {volume} {59}},\ \bibinfo {pages} {439} (\bibinfo {year} {2012})}\BibitemShut {NoStop}%
\bibitem [{\citenamefont {Farley}\ and\ \citenamefont {Wing}(1981)}]{farley1981}%
  \BibitemOpen
  \bibfield  {author} {\bibinfo {author} {\bibfnamefont {J.~W.}\ \bibnamefont {Farley}}\ and\ \bibinfo {author} {\bibfnamefont {W.~H.}\ \bibnamefont {Wing}},\ }\bibfield  {title} {\bibinfo {title} {Accurate calculation of dynamic stark shifts and depopulation rates of {R}ydberg energy levels induced by blackbody radiation. hydrogen, helium, and alkali-metal atoms},\ }\href {https://doi.org/10.1103/PhysRevA.23.2397} {\bibfield  {journal} {\bibinfo  {journal} {Phys. Rev. A}\ }\textbf {\bibinfo {volume} {23}},\ \bibinfo {pages} {2397} (\bibinfo {year} {1981})}\BibitemShut {NoStop}%
\bibitem [{\citenamefont {Fertig}\ and\ \citenamefont {Gibble}(2000)}]{gibble2000}%
  \BibitemOpen
  \bibfield  {author} {\bibinfo {author} {\bibfnamefont {C.}~\bibnamefont {Fertig}}\ and\ \bibinfo {author} {\bibfnamefont {K.}~\bibnamefont {Gibble}},\ }\bibfield  {title} {\bibinfo {title} {Measurement and cancellation of the cold collision frequency shift in an ${}^{87}\mathrm{Rb}$ fountain clock},\ }\href {https://doi.org/10.1103/PhysRevLett.85.1622} {\bibfield  {journal} {\bibinfo  {journal} {Phys. Rev. Lett.}\ }\textbf {\bibinfo {volume} {85}},\ \bibinfo {pages} {1622} (\bibinfo {year} {2000})}\BibitemShut {NoStop}%
\bibitem [{\citenamefont {Sortais}\ \emph {et~al.}(2000)\citenamefont {Sortais}, \citenamefont {Bize}, \citenamefont {Nicolas}, \citenamefont {Clairon}, \citenamefont {Salomon},\ and\ \citenamefont {Williams}}]{sortais2000}%
  \BibitemOpen
  \bibfield  {author} {\bibinfo {author} {\bibfnamefont {Y.}~\bibnamefont {Sortais}}, \bibinfo {author} {\bibfnamefont {S.}~\bibnamefont {Bize}}, \bibinfo {author} {\bibfnamefont {C.}~\bibnamefont {Nicolas}}, \bibinfo {author} {\bibfnamefont {A.}~\bibnamefont {Clairon}}, \bibinfo {author} {\bibfnamefont {C.}~\bibnamefont {Salomon}},\ and\ \bibinfo {author} {\bibfnamefont {C.}~\bibnamefont {Williams}},\ }\bibfield  {title} {\bibinfo {title} {Cold collision frequency shifts in a ${}^{87}\mathrm{Rb}$ atomic fountain},\ }\href {https://doi.org/10.1103/PhysRevLett.85.3117} {\bibfield  {journal} {\bibinfo  {journal} {Phys. Rev. Lett.}\ }\textbf {\bibinfo {volume} {85}},\ \bibinfo {pages} {3117} (\bibinfo {year} {2000})}\BibitemShut {NoStop}%
\bibitem [{\citenamefont {Dieckmann}\ \emph {et~al.}(1998)\citenamefont {Dieckmann}, \citenamefont {Spreeuw}, \citenamefont {Weidem\"uller},\ and\ \citenamefont {Walraven}}]{Dieckmann1998}%
  \BibitemOpen
  \bibfield  {author} {\bibinfo {author} {\bibfnamefont {K.}~\bibnamefont {Dieckmann}}, \bibinfo {author} {\bibfnamefont {R.~J.~C.}\ \bibnamefont {Spreeuw}}, \bibinfo {author} {\bibfnamefont {M.}~\bibnamefont {Weidem\"uller}},\ and\ \bibinfo {author} {\bibfnamefont {J.~T.~M.}\ \bibnamefont {Walraven}},\ }\bibfield  {title} {\bibinfo {title} {Two-dimensional magneto-optical trap as a source of slow atoms},\ }\href {https://doi.org/10.1103/PhysRevA.58.3891} {\bibfield  {journal} {\bibinfo  {journal} {Phys. Rev. A}\ }\textbf {\bibinfo {volume} {58}},\ \bibinfo {pages} {3891} (\bibinfo {year} {1998})}\BibitemShut {NoStop}%
\bibitem [{\citenamefont {Camposeo}\ \emph {et~al.}(2001)\citenamefont {Camposeo}, \citenamefont {Piombini}, \citenamefont {Cervelli}, \citenamefont {Tantussi}, \citenamefont {Fuso},\ and\ \citenamefont {Arimondo}}]{Camposeo2001}%
  \BibitemOpen
  \bibfield  {author} {\bibinfo {author} {\bibfnamefont {A.}~\bibnamefont {Camposeo}}, \bibinfo {author} {\bibfnamefont {A.}~\bibnamefont {Piombini}}, \bibinfo {author} {\bibfnamefont {F.}~\bibnamefont {Cervelli}}, \bibinfo {author} {\bibfnamefont {F.}~\bibnamefont {Tantussi}}, \bibinfo {author} {\bibfnamefont {F.}~\bibnamefont {Fuso}},\ and\ \bibinfo {author} {\bibfnamefont {E.}~\bibnamefont {Arimondo}},\ }\bibfield  {title} {\bibinfo {title} {A cold cesium atomic beam produced out of a pyramidal funnel},\ }\href {https://doi.org/https://doi.org/10.1016/S0030-4018(01)01643-1} {\bibfield  {journal} {\bibinfo  {journal} {Optics Communications}\ }\textbf {\bibinfo {volume} {200}},\ \bibinfo {pages} {231} (\bibinfo {year} {2001})}\BibitemShut {NoStop}%
\bibitem [{\citenamefont {Olson}\ \emph {et~al.}(2006)\citenamefont {Olson}, \citenamefont {Mhaskar},\ and\ \citenamefont {Raithel}}]{Olson2006}%
  \BibitemOpen
  \bibfield  {author} {\bibinfo {author} {\bibfnamefont {S.~E.}\ \bibnamefont {Olson}}, \bibinfo {author} {\bibfnamefont {R.~R.}\ \bibnamefont {Mhaskar}},\ and\ \bibinfo {author} {\bibfnamefont {G.}~\bibnamefont {Raithel}},\ }\bibfield  {title} {\bibinfo {title} {Continuous propagation and energy filtering of a cold atomic beam in a long high-gradient magnetic atom guide},\ }\href {https://doi.org/10.1103/PhysRevA.73.033622} {\bibfield  {journal} {\bibinfo  {journal} {Phys. Rev. A}\ }\textbf {\bibinfo {volume} {73}},\ \bibinfo {pages} {033622} (\bibinfo {year} {2006})}\BibitemShut {NoStop}%
\bibitem [{\citenamefont {Chen}\ \emph {et~al.}(2014)\citenamefont {Chen}, \citenamefont {Zigo},\ and\ \citenamefont {Raithel}}]{Chen2014}%
  \BibitemOpen
  \bibfield  {author} {\bibinfo {author} {\bibfnamefont {Y.-J.}\ \bibnamefont {Chen}}, \bibinfo {author} {\bibfnamefont {S.}~\bibnamefont {Zigo}},\ and\ \bibinfo {author} {\bibfnamefont {G.}~\bibnamefont {Raithel}},\ }\bibfield  {title} {\bibinfo {title} {Atom trapping and spectroscopy in cavity-generated optical potentials},\ }\href {https://doi.org/10.1103/PhysRevA.89.063409} {\bibfield  {journal} {\bibinfo  {journal} {Phys. Rev. A}\ }\textbf {\bibinfo {volume} {89}},\ \bibinfo {pages} {063409} (\bibinfo {year} {2014})}\BibitemShut {NoStop}%
\bibitem [{\citenamefont {Strangfeld}\ \emph {et~al.}(2021)\citenamefont {Strangfeld}, \citenamefont {Kanthak}, \citenamefont {Schiemangk}, \citenamefont {Wiegand}, \citenamefont {Wicht}, \citenamefont {Ling},\ and\ \citenamefont {Krutzik}}]{Strangfeld2021}%
  \BibitemOpen
  \bibfield  {author} {\bibinfo {author} {\bibfnamefont {A.}~\bibnamefont {Strangfeld}}, \bibinfo {author} {\bibfnamefont {S.}~\bibnamefont {Kanthak}}, \bibinfo {author} {\bibfnamefont {M.}~\bibnamefont {Schiemangk}}, \bibinfo {author} {\bibfnamefont {B.}~\bibnamefont {Wiegand}}, \bibinfo {author} {\bibfnamefont {A.}~\bibnamefont {Wicht}}, \bibinfo {author} {\bibfnamefont {A.}~\bibnamefont {Ling}},\ and\ \bibinfo {author} {\bibfnamefont {M.}~\bibnamefont {Krutzik}},\ }\bibfield  {title} {\bibinfo {title} {Prototype of a compact rubidium-based optical frequency reference for operation on nanosatellites},\ }\href {https://doi.org/10.1364/JOSAB.420875} {\bibfield  {journal} {\bibinfo  {journal} {JOSA B}\ }\textbf {\bibinfo {volume} {38}},\ \bibinfo {pages} {1885} (\bibinfo {year} {2021})}\BibitemShut {NoStop}%
\bibitem [{\citenamefont {Ropp}\ \emph {et~al.}(2023)\citenamefont {Ropp}, \citenamefont {Zhu}, \citenamefont {Yulaev}, \citenamefont {Westly}, \citenamefont {Simelgor}, \citenamefont {Rakholia}, \citenamefont {Lunden}, \citenamefont {Sheredy}, \citenamefont {Boyd}, \citenamefont {Papp}, \citenamefont {Agrawal},\ and\ \citenamefont {Aksyuk}}]{ropp2023}%
  \BibitemOpen
  \bibfield  {author} {\bibinfo {author} {\bibfnamefont {C.}~\bibnamefont {Ropp}}, \bibinfo {author} {\bibfnamefont {W.}~\bibnamefont {Zhu}}, \bibinfo {author} {\bibfnamefont {A.}~\bibnamefont {Yulaev}}, \bibinfo {author} {\bibfnamefont {D.}~\bibnamefont {Westly}}, \bibinfo {author} {\bibfnamefont {G.}~\bibnamefont {Simelgor}}, \bibinfo {author} {\bibfnamefont {A.}~\bibnamefont {Rakholia}}, \bibinfo {author} {\bibfnamefont {W.}~\bibnamefont {Lunden}}, \bibinfo {author} {\bibfnamefont {D.}~\bibnamefont {Sheredy}}, \bibinfo {author} {\bibfnamefont {M.~M.}\ \bibnamefont {Boyd}}, \bibinfo {author} {\bibfnamefont {S.}~\bibnamefont {Papp}}, \bibinfo {author} {\bibfnamefont {A.}~\bibnamefont {Agrawal}},\ and\ \bibinfo {author} {\bibfnamefont {V.}~\bibnamefont {Aksyuk}},\ }\bibfield  {title} {\bibinfo {title} {Integrating planar photonics for multi-beam generation and atomic clock packaging on chip},\ }\href {https://doi.org/10.1038/s41377-023-01081-x} {\bibfield  {journal} {\bibinfo  {journal} {Light: Sci. Appl.}\
  }\textbf {\bibinfo {volume} {12}},\ \bibinfo {pages} {83} (\bibinfo {year} {2023})}\BibitemShut {NoStop}%
\bibitem [{\citenamefont {Tran~Tan}\ and\ \citenamefont {Derevianko}(2023)}]{trantan2023}%
  \BibitemOpen
  \bibfield  {author} {\bibinfo {author} {\bibfnamefont {H.~B.}\ \bibnamefont {Tran~Tan}}\ and\ \bibinfo {author} {\bibfnamefont {A.}~\bibnamefont {Derevianko}},\ }\bibfield  {title} {\bibinfo {title} {Precision theoretical determination of electric-dipole matrix elements in atomic cesium},\ }\href {https://doi.org/10.1103/PhysRevA.107.042809} {\bibfield  {journal} {\bibinfo  {journal} {Phys. Rev. A}\ }\textbf {\bibinfo {volume} {107}},\ \bibinfo {pages} {042809} (\bibinfo {year} {2023})}\BibitemShut {NoStop}%
\end{thebibliography}%

\end{document}